\documentclass[journal=jctc,manuscript=article]{achemso}
\setkeys{acs}{maxauthors=0,articletitle=true}

\usepackage{achemso}
\usepackage{graphics}
\usepackage{amssymb,amsfonts}
\usepackage{graphicx}
\usepackage[table]{xcolor}
\usepackage{multirow}
\usepackage{caption}
\usepackage{subcaption}
\usepackage{booktabs}
\usepackage{colortbl}
\usepackage{amsmath}
\usepackage{amsopn}
\usepackage{siunitx}
\usepackage{bm}
\usepackage{color}
\usepackage{array}
\usepackage{lscape}
\usepackage{mciteplus}
\usepackage[version=3]{mhchem}
\usepackage{ulem}
\usepackage{listings}
\usepackage{enumerate}

\SectionNumbersOn

\author{Janus J. Eriksen}
\email{janus.eriksen@bristol.ac.uk}
\affiliation[Johannes Gutenberg-Universit\"at Mainz]
{Institut f\"ur Physikalische Chemie, Johannes Gutenberg-Universit\"at Mainz, Duesbergweg 10-14, 55128 Mainz, Germany}
\altaffiliation{Present address: {\it{School of Chemistry, University of Bristol, Cantock's Close, Bristol BS8 1TS, United Kingdom}}}
\author{J{\"u}rgen Gauss}
\email{gauss@uni-mainz.de}
\affiliation[Johannes Gutenberg-Universit\"at Mainz]
{Institut f\"ur Physikalische Chemie, Johannes Gutenberg-Universit\"at Mainz, Duesbergweg 10-14, 55128 Mainz, Germany}

\title[TITLE]{Many-Body Expanded Full Configuration Interaction. II. Strongly Correlated Regime}

\begin{document}

%
%
\begin{abstract}

In this second part of our series on the recently proposed many-body expanded full configuration interaction (MBE-FCI) method, we introduce the concept of multideterminantal expansion references. Through theoretical arguments and numerical validations, the use of this class of starting points is shown to result in a focussed compression of the MBE decomposition of the FCI energy, thus allowing chemical problems dominated by strong correlation to be addressed by the method. The general applicability and performance enhancements of MBE-FCI are verified for standard stress tests such as the bond dissociations in H$_2$O, N$_2$, C$_2$, and a linear H$_{10}$ chain. Furthermore, the benefits of employing a multideterminantal expansion reference in accelerating calculations of high accuracy are discussed, with an emphasis on calculations in extended basis sets. As an illustration of this latter quality of the MBE-FCI method, results for H$_2$O and C$_2$ in basis sets ranging from double- to pentuple-$\zeta$ quality are presented, demonstrating near-ideal parallel scaling on up to almost $25000$ processing units.

\end{abstract}

\newpage

%
%

\section{Introduction}\label{intro_section}

Leveraged by the technological progress of high-performance scientific computing, in combination with a continuous decrease in relative cost and increase in general availability of suitable hardware resources, the field of quantum chemistry has flourished notably over the past half-century. These days, it has even matured to such a predictive state where one may often describe microscopic events in molecules and matter by means of computers at a level on par with what is achievable experimentally. As an illustrative example of such advancements, the early idea of selected configuration interaction~\cite{malrieu_selected_ci_jcp_1973} as a procedure to perform a focused sampling of the exact full configuration interaction~\cite{knowles_handy_fci_cpl_1984,olsen_fci_jcp_1988,olsen_fci_cpl_1990} (FCI) wave function has seen a remarkable revival in the literature in recent years~\cite{booth_alavi_fciqmc_jcp_2009,cleland_booth_alavi_jcp_2010,booth_alavi_fciqmc_nature_2013,petruzielo_umrigar_spmc_prl_2012,holmes_umrigar_heat_bath_fock_space_jctc_2016,holmes_umrigar_heat_bath_ci_jctc_2016,sharma_umrigar_heat_bath_ci_jctc_2017,li_sharma_umrigar_heat_bath_ci_jcp_2018,tubman_whaley_selected_ci_jcp_2016,schriber_evangelista_selected_ci_jcp_2016,schriber_evangelista_adaptive_ci_jctc_2017,liu_hoffmann_ici_jctc_2016,loos_selected_ci_jcp_2018,loos_jacquemin_selected_ci_exc_state_jctc_2018,fales_koch_martinez_rrfci_jctc_2018,coe_ml_ci_jctc_2018}. While all of the different emerging incarnations of the theory differ slightly in their algorithmic and implementational details, the common denominator remains the quest for a quantitative approximation to the $N$-dimensional FCI wave function ($N$ being the number of electrons). Likewise, methods which, rather than sampling the wave function for individual important contributions, instead target the associated FCI energy directly without making reference to individual determinants have also been relaunched~\cite{zimmerman_ifci_jcp_2017_1,zimmerman_ifci_jcp_2017_2,eriksen_mbe_fci_jpcl_2017,eriksen_mbe_fci_weak_corr_jctc_2018}. For example, the recently proposed many-body expanded FCI~\cite{eriksen_mbe_fci_jpcl_2017} (MBE-FCI) method accomplishes this via a many-body expansion (MBE) in a basis of the virtual molecular orbitals (MOs) of a preceding Hartree-Fock (HF) calculation. Facilitated by a screening protocol to ensure rapid convergence to the exact FCI target, and aided as well as accelerated by intermediate coupled cluster (CC) base models, the initial set of proof-of-concept results for the MBE-FCI method was augmented in the first part of the present series (Ref. \citenum{eriksen_mbe_fci_weak_corr_jctc_2018}) by results for a selection of prototypical single-reference, weakly correlated molecular systems.\\

\begin{figure}[ht]
\begin{center}
\includegraphics{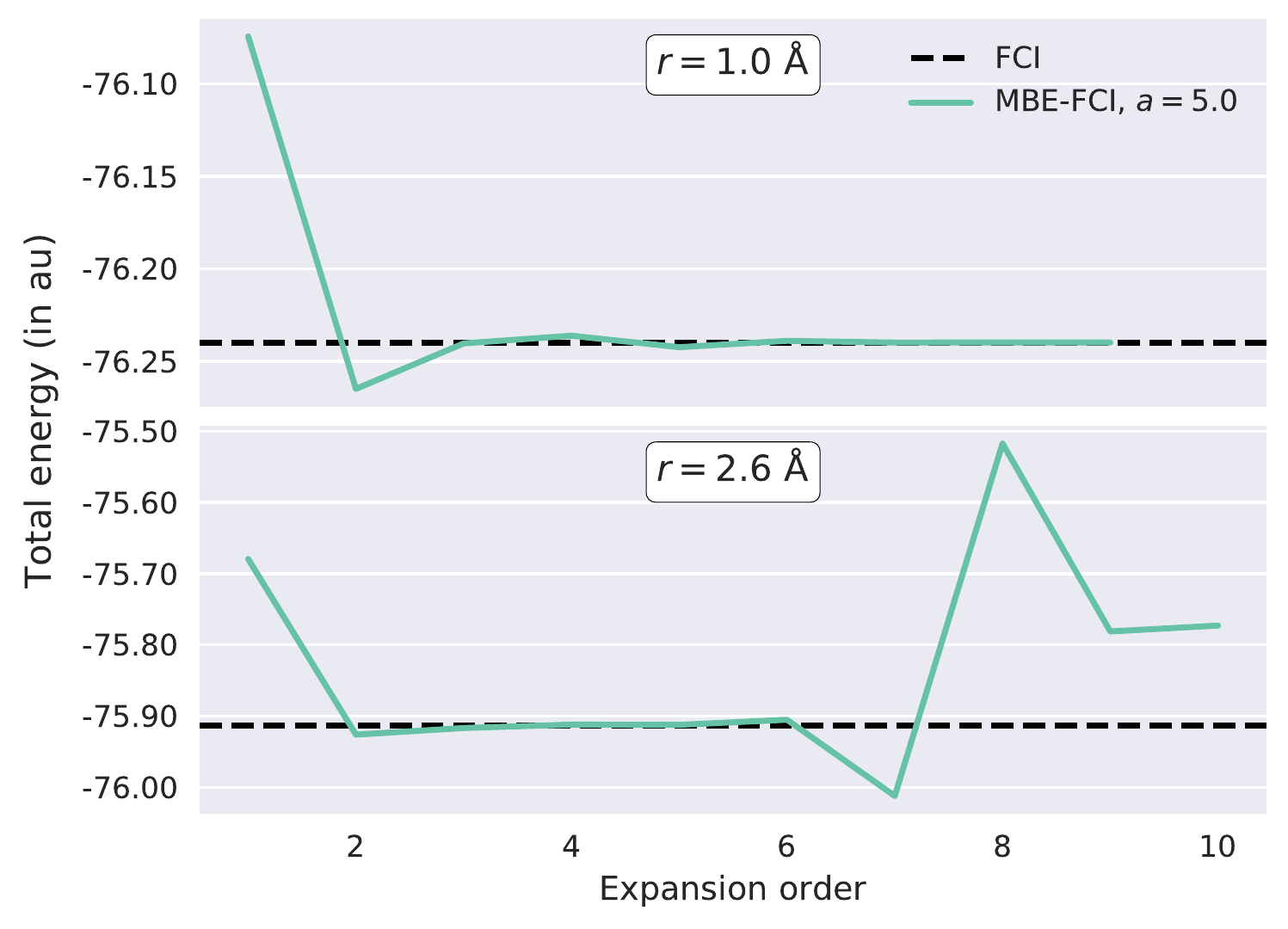}
\caption{Energy convergence of MBE-FCI expansions (no base model, screening threshold of $a = 5.0$) for the ${^{1}}A_{1}$ ground state of H$_2$O in a cc-pVDZ basis set~\cite{dunning_1_orig} at bond distances $r = 1.0$ {\AA} and $r = 2.6$ {\AA} during the symmetric bond stretch ($\angle 104.2^\circ$, $C_{2\text{v}}$ symmetry).}
\label{h2o_dz_problem_1_fig}
\end{center}
\end{figure}
Despite the reported accuracy with respect to FCI, one critical shortcoming of the current version of the MBE-FCI method was commented on in the closing part of Ref. \citenum{eriksen_mbe_fci_weak_corr_jctc_2018}, namely its application to systems dominated by strong electron correlation. As a demonstration of this, Figure \ref{h2o_dz_problem_1_fig} shows the convergence of MBE-FCI calculations on the ground state of water near equilibrium and at a stretched geometry. Given the simplicity of the methodology, the only plausible cause of the deterioration of the method at the latter of these two geometries is related to the inadequacy of a single Slater determinant acting as the reference for the expansion. Thus, any avenue toward a functional MBE-FCI method for strongly correlated systems must accordingly, in one way or another, seek to eliminate this HF dependency.\\

In the present work, we are proposing a generalization of the MBE-FCI method to arbitrary multideterminantal expansion references such as those of complete active space (CAS) self-consistent field~\cite{roos_casscf_acp_1987} (CASSCF) or configuration interaction (CASCI) theory, depending on whether or not orbital relaxation is accounted for or not. Common for both of these references is the fact that they are tailored to recover the correlation associated with potential degeneracies present in a chosen active reference space. However, they will still need to be corrected for correlation effects outside of this. In the following, we will show how to formulate MBE-FCI on top of any of these two references in order to compute near-exact electronic energies in both the weakly and the strongly correlated regime. In combination with the parallel computing potential of the method, this feature enhancement of MBE-FCI will allow for precise results to be obtained for chemical problems of any nature in extended basis sets. In fact, we will demonstrate that the use of a multideterminantal expansion reference generally results in a focussed compression of the involved MBE, in the sense that a large number of contributions to the expansions will be prescreened and thus deliberately not accounted for in the decomposition of the FCI correlation energy. As such, we will show that appropriately chosen expansion references may even result in higher accuracy for calculations in basis sets of unprecedented size. While MBE-FCI results may be readily obtained on commodity hardware, we will end by reporting and discussing near-ideal scaling results on close to $25000$ physical cores.

%
%

\section{Theory}\label{theory_section}

As outlined in Ref. \citenum{eriksen_mbe_fci_weak_corr_jctc_2018}, the master equation behind the MBE-FCI method is the following decomposition of the FCI correlation energy formulated in terms of virtual spatial MOs (conventionally labelled by indices $a,b,c,\ldots$)
\begin{align}
E_{\text{FCI}} &= \sum_{a}\epsilon_{a} + \sum_{a<b}\Delta\epsilon_{ab} + \sum_{a<b<c}\Delta\epsilon_{abc} + \ldots \nonumber \\
&\equiv E^{(1)} + E^{(2)} + E^{(3)} + \ldots + E^{(M_{\text{v}})} \label{mbe_eq}
\end{align}
where $M_{\text{v}}$ denotes the number of virtual MOs in the system and $\epsilon_{a}$ is the energy of a CASCI calculation in the composite space of virtual orbital $a$ and the complete set of occupied orbitals. The increments of order $n$, $\Delta\epsilon_{[\Omega]_{n}}$, which account for changes in the correlation energy from allowing for electronic excitations into $n$ over $n-1$ virtual orbitals, are recursively defined through the following relation for a general tuple of $n$ virtual MOs, $[\Omega]_{n} \equiv [abc\cdots]_{n}$, as
\begin{align}
\Delta\epsilon_{[\Omega]_{n}} = \epsilon_{[\Omega]_{n}} - \sum_{p \in S_1[\Omega]_{n}}\epsilon_{p} - \sum_{pq \in S_2[\Omega]_{n}}\Delta\epsilon_{pq} - \ldots - \sum_{pqrs\cdots \in S_{n-1}[\Omega]_{n}}\Delta\epsilon_{pqrs\cdots} \label{increment_eq}
\end{align}
In Eq. \ref{increment_eq}, the action of $S_{m}$ onto $[\Omega]_{n}$ is to construct all possible unique subtuples of order (length) $m$ where $1\leq m<n$.\\

\begin{figure}[ht]
\begin{center}
\includegraphics[width=\textwidth,bb=25 21 999 763]{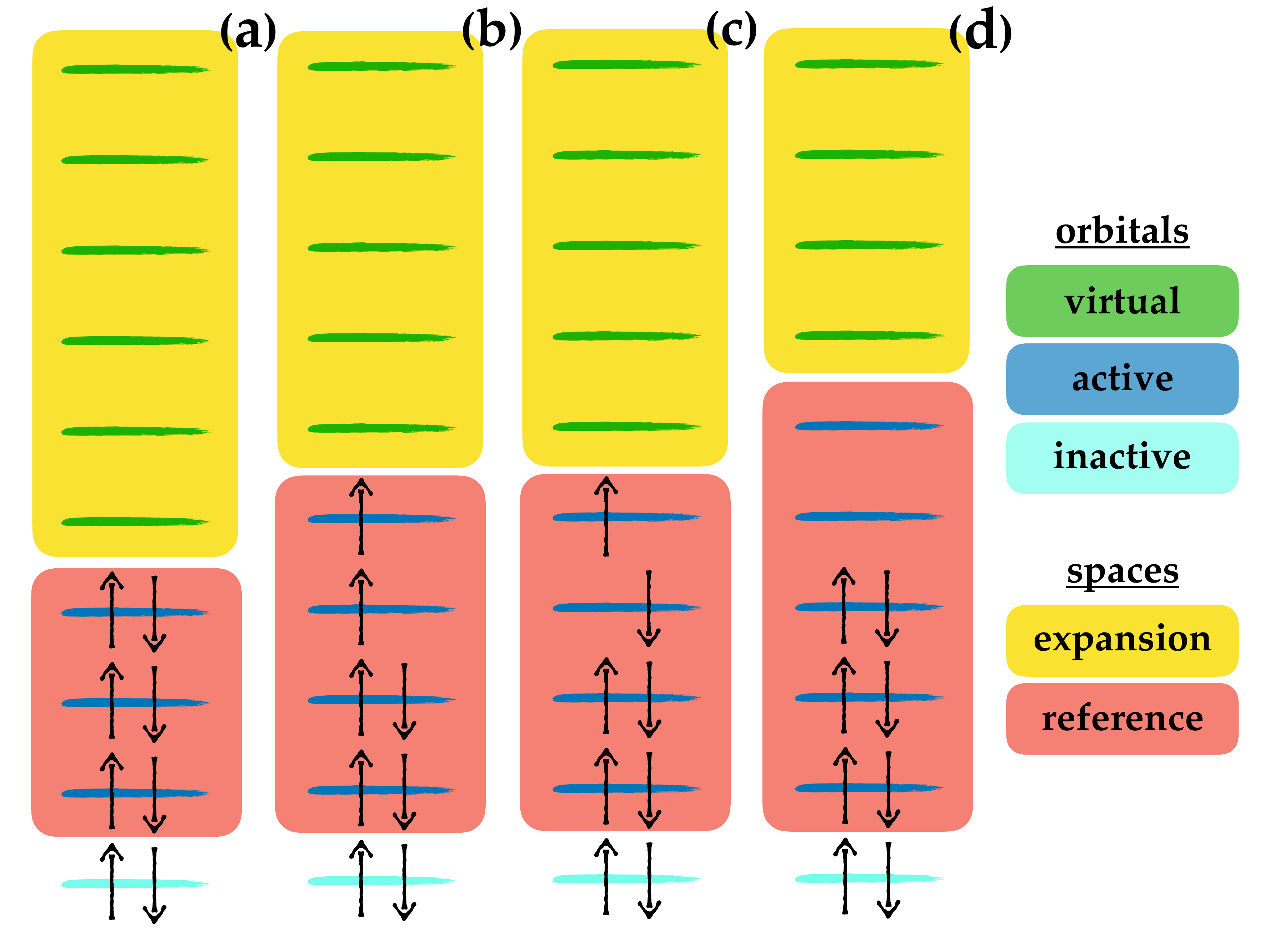}
\caption{Schematic representation of the various manners in which to treat a general ($6,9$) system using the MBE-FCI method. The most dominant determinant for each case is indicated by the distribution of $\alpha$- and $\beta$-electrons (case {\textbf{(c)}} is degenerate). In all four cases, 2 frozen (inactive) electrons in a single core orbital are excluded from the MBE-FCI calculation.}
\label{mo_diagram_fig}
\end{center}
\end{figure}
By inspecting Eq. \ref{mbe_eq}, it is evident that the MBE-FCI method---as described above---will be biased toward the closed-shell HF reference used to partition the complete set of MOs into a reference space, which comprises all occupied MOs, and an expansion space defined by the complete set of virtual MOs. This partitioning is depicted as case {\textbf{(a)}} in Figure \ref{mo_diagram_fig} for a general system of 6 electrons in 9 orbitals (or, in short-hand notation, a ($6,9$) system). For this specific case, we assume that the restricted HF (RHF) determinant is indeed comprising a fair approximation to the FCI wave function, i.e., it has the dominant weight. The MBE-FCI decomposition of the FCI correlation energy will involve a total of $6$ CASCI($6,4$) calculations at order $1$ (the ($6,3$) RHF reference space augmented by all possible single virtual MOs of the expansion space), $15$ CASCI($6,5$) calculations at order $2$ (augmentation by all unique pair combinations of virtual MOs), etc., and culminate in a single CASCI(6,9) calculation at order $6$ (i.e., FCI for system {\textbf{(a)}}). The corresponding partitioning for a high-spin open-shell triplet with $[4,2]$ electrons of $[\alpha,\beta]$ spin is depicted as case {\textbf{(b)}} in Figure \ref{mo_diagram_fig}. Here, the ($[4,2],4$) restricted open-shell HF (ROHF) determinant is assumed dominant. For a system like this, the correlation energy of a CASCI($[4,2],4$) calculation in the reference space may be non-zero, unlike in case {\textbf{(a)}}, so this zeroth-order calculation must precede the actual expansion which otherwise proceeds as above, albeit in terms of a reduced expansion space. At order $1$, a total of $5$ unique CASCI($[4,2],5$) calculations are performed, followed by $10$ CASCI($[4,2],6$) calculations at order $2$, etc., ending with a single CASCI($[4,2],9$) calculation at order $5$ (i.e., FCI for system {\textbf{(b)}}).\\

Now, instead of a dominant RHF determinant, let us rather assume that the FCI wave function is comprised of a number of determinants with large coefficients. As an example of this, case {\textbf{(c)}} in Figure \ref{mo_diagram_fig} depicts the non-trivial case of a pair of degenerate open-shell singlet determinants carrying the largest weights. In this case, the reference space will hence need to encompass all important closed- and open-shell singlets. Similarly to the open-shell case {\textbf{(b)}}, a CASCI($6,4$) reference space calculation then precedes the actual MBE-FCI expansion. This is otherwise initiated at order $1$ with $5$ CASCI($6,5$) calculations (each involving the reference space and a single virtual MO of the expansion space) and ends in a single CASCI($6,9$) calculation at order $5$ (i.e., FCI for system {\textbf{(c)}}). In analogy with case {\textbf{(c)}}, one may similarly choose to expand the reference space even further. For instance, for the weakly correlated system case {\textbf{(a)}}, we may extend the native RHF reference space by including, say, a pair of virtual MOs. This is depicted as case {\textbf{(d)}} in Figure \ref{mo_diagram_fig} with the LUMO and LUMO+1 included in the reference space. In comparison with the MBE-FCI expansion for case {\textbf{(a)}}, this choice of reference corresponds to excluding all incremental terms to the FCI decomposition that fail to make reference to this given orbital pair. While the final result---in the absence of any screening---will be the same for cases {\textbf{(a)}} and {\textbf{(d)}}, the latter will converge in terms of a reduced number of individual CASCI calculations.\\

In general terms, this flexibility of the MBE may in principle allow for any point of initiation in the treatment of electron correlation. For multireference cases, it allows for dedicating separate attention to a chemically motivated active space by singling out the MOs belonging to this and have MBE-FCI account for all dynamic correlation out of the space (as in case {\textbf{(c)}} of Figure \ref{mo_diagram_fig}). Phrased slightly differently, whenever the reference space is multideterminantal, the individual CASCI calculations of an MBE-FCI expansion always correlate the MOs of the active space in addition to an increasing number of virtual MOs. Denoting the reference space as $[\Pi]$, the expression in Eq. \ref{increment_eq} for an $n$th-order increment may be generalized as
\begin{align}
\Delta\epsilon_{[\Omega]_{n}\supset[\Pi]} = \epsilon_{[\Omega]_{n}\supset[\Pi]} - \epsilon_{[\Pi]} - \ldots - \sum_{pqrs\cdots \in S_{n-1}[\Omega]_{n}\supset[\Pi]}\Delta\epsilon_{pqrs\cdots} \label{increment_as_eq}
\end{align}
where every orbital tuple entering Eq. \ref{increment_as_eq} is required to form a superset of the reference space. We note the universality of Eq. \ref{increment_as_eq} as no changes with respect to the standard formulation of MBE-FCI are introduced whenever the reference space coincides with the HF determinant~\cite{eriksen_mbe_fci_jpcl_2017,eriksen_mbe_fci_weak_corr_jctc_2018}.\\

By virtue of Eq. \ref{increment_as_eq}, one is hence left with two distinct choices for the {\it{expansion reference}}: CASCI or CASSCF, which differ in terms of what orbitals are used as expansion objects. In this notation, the closed- and open-shell calculations of Ref. \citenum{eriksen_mbe_fci_weak_corr_jctc_2018} employed minimal CASCI references that encompass only the HF determinant, as in cases {\textbf{(a)}} and {\textbf{(b)}} of Figure \ref{mo_diagram_fig}. When the reference is further spanned by the canonical MOs of a preceding RHF/ROHF calculation, we will denote this as an RHF reference space. Furthermore, we note how the use of a multideterminantal expansion reference does not preclude the use of any of the expansion bases of Ref. \citenum{eriksen_mbe_fci_weak_corr_jctc_2018}. However, these are bound to perform inadequately in the multireference regime as they are intended for capturing weak correlation.\\

Finally, due to the freedom to employ multideterminantal expansion references, some of the black-box nature of the MBE-FCI method might appear lost. In the present study, also for reasons related to the reproducibility of the results to follow in Section \ref{results_section}, we will restrict ourselves to simple expansion references defined in terms of the symmetries of the underlying orbital spaces. However, we are currently working on selection schemes specific to MBE-FCI which automatically choose an optimal expansion reference for a given chemical problem. As this remains work in progress, its details are postponed to later stages of this series.

%
%

\section{Computational Details}\label{comp_section}

The code used to perform the MBE-FCI calculations of the present work is the {\textsc{pymbe}} code~\cite{pymbe}, which is written in Python/NumPy~\cite{numpy} and utilizes the {\textsc{pyscf}} program~\cite{pyscf_paper,pyscf_prog} for all electronic structure kernels. As discussed in Ref. \citenum{eriksen_mbe_fci_weak_corr_jctc_2018}, the {\textsc{pymbe}} code has been parallelized by means of the implementation of the message passing interface (MPI) standard in the {\sc{mpi4py}} Python module~\cite{mpi4py_1,mpi4py_2,mpi4py_3} and supports full Abelian point-group symmetry. However, for the linear diatomics (N$_2$ and C$_2$) of Sections \ref{n2_subsubsection} and \ref{c2_subsubsection}, one may not distinguish between certain states based on symmetry arguments alone as, e.g., ${^{1}}\Sigma^{+}_g$ and ${^{1}}\Delta_g$ states will be spanned by the same irreducible representation ($A_g$) in the $D_{2\text{h}}$ subgroup of the true $D_{\infty\text{h}}$ symmetry group. For that reason, we have implemented a prescreening filter which works to alleviate this inconsistency by only calculating increments for which both the $x$- and $y$-component of a given pair of degenerate $\pi_u$- or $\pi_g$-orbitals are simultaneously included in the respective CASCI calculation. We denote this filtering as $\pi$-{\it{pruning}}, in the present case aimed toward ${^{1}}\Sigma^{+}_g$ states, but corresponding pruning schemes for focussing on, e.g., ${^{1}}\Delta_g$ states may be implemented in an analogous manner. Effectively, the use of $\pi$-pruning is shown for C$_2$ in Section \ref{c2_subsubsection} to be crucial near crossings of ${^{1}}\Sigma^{+}_g$ and ${^{1}}\Delta_g$ states and we will further demonstrate how its use generally results in a drastically reduced number of involved CASCI calculations, without sacrificing the overall accuracy of the MBE-FCI method.\\

All of the results to follow in Section \ref{results_section} have been obtained on either of two computational resources, one local (i) and one external (ii): (i) a single Intel Xeon Broadwell E$5$--$2699$ $\text{v}4$ node with a total of $44$ cores {@} $2.20$ GHz and $768$ GB of global memory, and (ii) the Hazel Hen system at HLRS, Universit{\"a}t Stuttgart, which is a Cray $\text{XC}40$ supercomputer equipped with $7712$ Intel Xeon E$5$--$2680$ $\text{v}3$ nodes, each comprising $24$ cores {@} $2.5$ GHz and $128$ GB of global memory. In Section \ref{parallel_scaling_subsubsection}, we present results obtained on Hazel Hen illustrating the intra- and internode parallel performance of {\textsc{pymbe}}.\\

In general, the frozen-core approximation has been invoked throughout except in the calculations of the potential energy curve (PEC) of the linear H$_{10}$ chain in Section \ref{h10_subsubsection}. All reference data have been calculated using the {\sc{cfour}} quantum chemical program package~\cite{cfour,lipparini_gauss_rel_casscf_jctc_2016,ncc}. 

%
%

\section{Results}\label{results_section}

In the present section, we will report MBE-FCI/cc-pVDZ results for the PECs of H$_2$O, N$_2$, C$_2$, and H$_{10}$ in Section \ref{strong_corr_subsection}, while single-point calculations in more extended basis sets are presented in Section \ref{h2o_c2_equil_subsection}. Tabulated data are collected in the Supporting Information (SI). For the linear diatomics in Sections \ref{n2_subsubsection} and \ref{c2_subsubsection}, these are treated using $D_{2\text{h}}$ symmetry with $\pi$-pruning enabled if not otherwise noted, cf. Section \ref{comp_section}. For clarity, however, we will adhere to $D_{\infty\text{h}}$ term symbols for designating the involved electronic states.

\subsection{Strongly Correlated Systems}\label{strong_corr_subsection}

In the following, we will compare MBE-FCI results against high-level CC with up to quadruple excitations~\cite{ccsdtq_paper_1_jcp_1991,ccsdtq_paper_2_jcp_1992} (CCSDTQ) for all systems, FCI reference data wherever applicable (H$_2$O, N$_2$, and C$_2$ in Sections \ref{h2o_stretch_subsubsection} through \ref{c2_subsubsection}), FCI quantum Monte Carlo in its initiator adaption~\cite{booth_alavi_fciqmc_jcp_2009,cleland_booth_alavi_jcp_2010} ($i$-FCIQMC) for C$_2$ in Section \ref{c2_subsubsection}, and both internally contracted truncated multireference configuration interaction~\cite{mrci_werner_knowles_jcp_1988,mrci_werner_knowles_cpl_1988} (MRCI+Q) and density matrix renormalization group~\cite{white_dmrg_prl_1992,white_dmrg_prb_1993,white_martin_dmrg_jcp_1999,chan_head_gordon_dmrg_jcp_2002} (DMRG) theory in the case of the H$_{10}$ system in Section \ref{h10_subsubsection}.

\subsubsection{Symmetric Stretching of Water}\label{h2o_stretch_subsubsection}

\begin{figure}[ht]
\begin{center}
\includegraphics{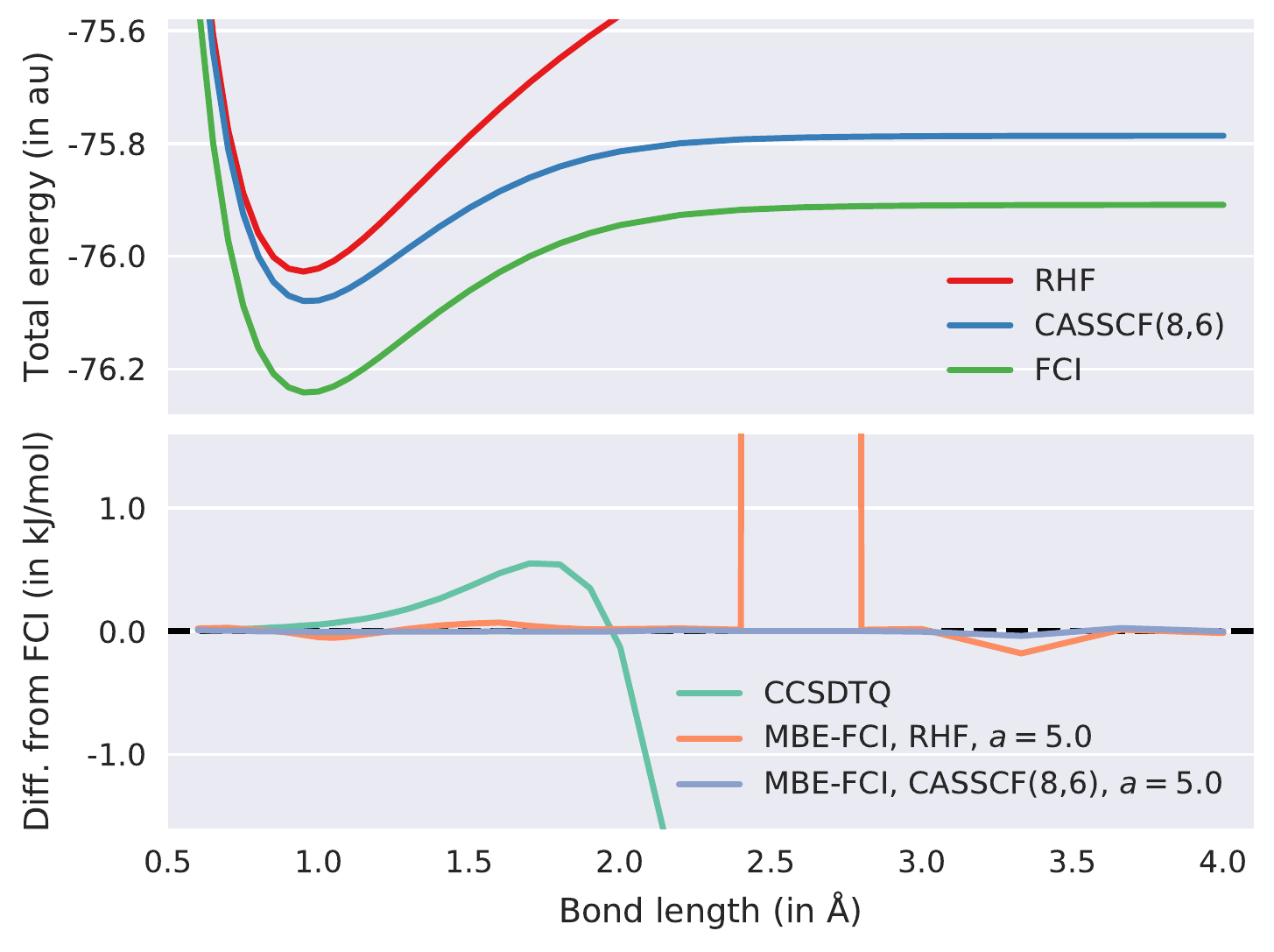}
\caption{Upper panel: RHF, CASSCF($8,6$), and FCI results for the ${^{1}}A_{1}$ ground state of H$_2$O in a cc-pVDZ basis set ($\angle 104.2^\circ$, $C_{2\text{v}}$ symmetry). Lower panel: Deviation of the CCSDTQ and MBE-FCI results from FCI.}
\label{h2o_dz_res_fig}
\end{center}
\end{figure}

As meticulously studied in the literature, for instance by Olsen {\it{et al.}} more than two decades ago~\cite{olsen_bond_break_h2o_jcp_1996}, the symmetric stretch of the two O--H bonds in water induces a slow, yet steady increase of static correlation. As such, the exercise of reproducing the FCI binding curve by means of approximate single- and multireference methods has been the topic of numerous studies~\cite{chan_bond_break_h2o_jcp_2003,krakauer_bond_break_h2o_jcp_2006,taube_bartlett_jcp_1_2008}. While at equilibrium, the FCI wave function for the ${^{1}}A_{1}$ ground state is almost entirely dominated by the RHF determinant, the wave function eventually goes on to describe a total of four unpaired electrons (product of O(${^{3}}P$) and two H(${^{2}}S$) states) upon approaching dissociation, as evidenced by the occupation numbers of the involved orbitals~\cite{olsen_bond_break_h2o_jcp_1996}. Hence, no single-reference model formulated in terms of the RHF reference---e.g., one of the CC hierarchy---may possibly be capable of offering a quantitative description throughout the entire domain mapped by Figure \ref{h2o_dz_res_fig}. As an example of this, we note that the CCSDTQ solution collapses at a bond elongation of about $200\%$, cf. the lower panel of Figure \ref{h2o_dz_res_fig} which displays the total deviation from the FCI reference results. A simple valence space CASSCF($8,6$) solution, on the other hand, gives a qualitatively correct description of the dissociation process in comparison with FCI, although the neglected dynamic correlation is obviously not negligible.\\

\begin{figure}[ht]
\begin{center}
\includegraphics{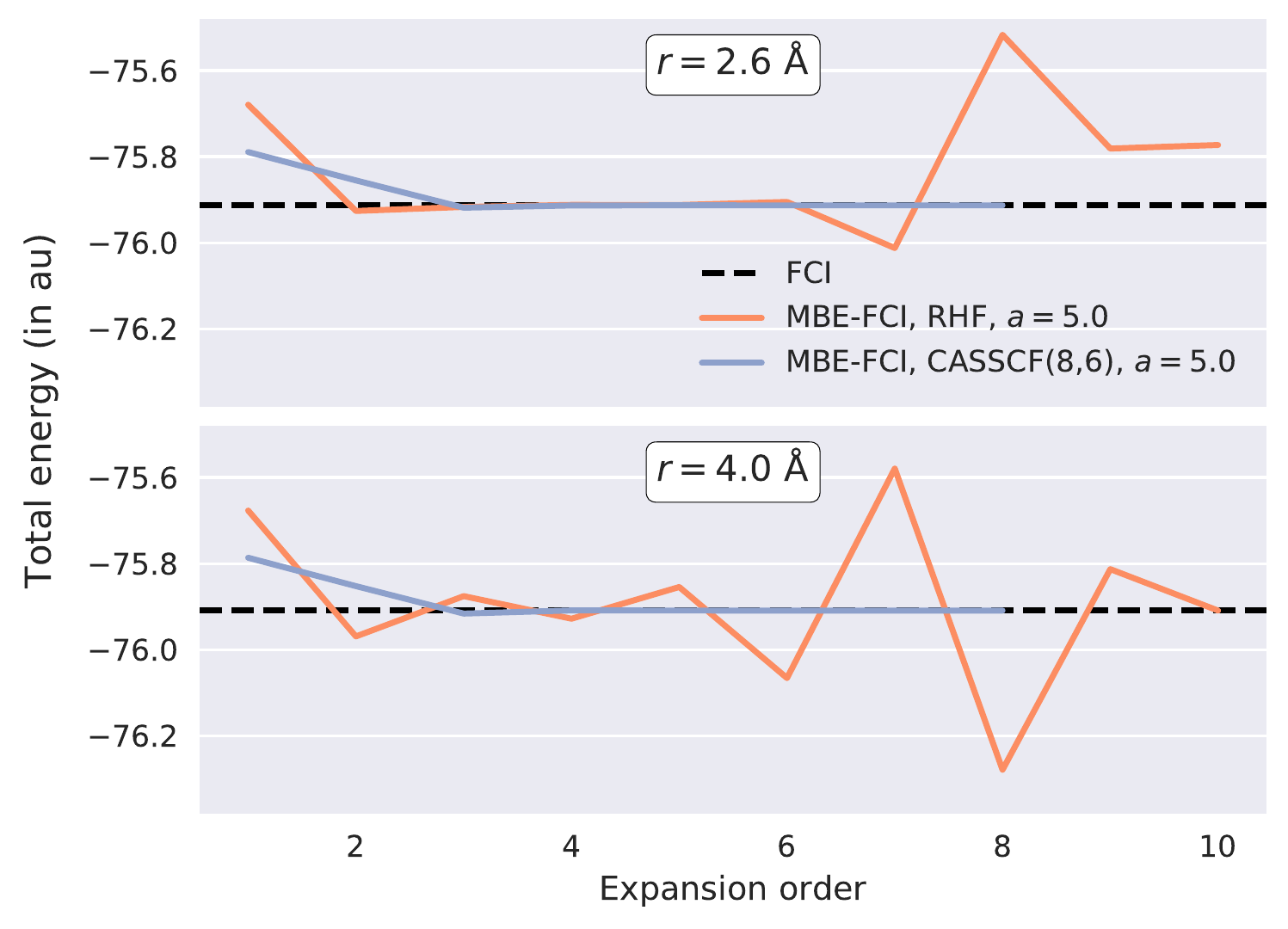}
\caption{Convergence of MBE-FCI expansions at O--H bond distances $r = 2.6$ {\AA} and $r = 4.0$ {\AA}.}
\label{h2o_dz_problem_2_fig}
\end{center}
\end{figure}

Figure \ref{h2o_dz_res_fig} furthermore presents the corresponding MBE-FCI deviations from FCI, using either the RHF or CASSCF($8,6$) solution as expansion reference. Commenting first on the results obtained with the latter of the two references, the deviations are observed to be uniform for all considered O--H bond lengths and thus independent of the nature of the correlation in H$_2$O. Bearing in mind that the employed screening threshold ($a=5.0$) is generally considered somewhat aggressive in the absence of a base model~\cite{eriksen_mbe_fci_weak_corr_jctc_2018}, this performance is convincing, albeit perhaps also serving as an indication of the static correlation present in stretched H$_2$O not being as `strong' as for some of the examples that will follow. On the contrary, the quality of the results for the MBE-FCI expansion starting from the RHF determinant is observed to deteriorate when moving toward the dissociation limit. In particular, a pronounced irregularity in the results is observed at a distance of $r = 2.6$ {\AA} (see also Figure \ref{h2o_dz_problem_1_fig}). This raises the question of what, if anything, is peculiar about this particular bond length. To shed light on this, the convergence profiles of the two MBE-FCI expansions of Figure \ref{h2o_dz_res_fig} for two selected stretched O--H bond distances ($r = 2.6$ {\AA} and $r = 4.0$ {\AA}) are compared in Figure \ref{h2o_dz_problem_2_fig}.\\

As is evident from Figure \ref{h2o_dz_problem_2_fig}, the convergence (with respect to our screening protocol) of the MBE-FCI expansion with an RHF reference is in fact significantly more erratic at the longer of the two distances. We note that this is not a unique feature of the two highlighted distances, but rather a commonly observed trend. Thus, the overall close agreement of these MBE-FCI results in Figure \ref{h2o_dz_res_fig} are in fact fortuitous at best. On the other hand, the MBE-FCI expansion formulated on top of the CASSCF reference is seen to yield the right results for the right reason.\\

In fact, in terms of the amount of individual CASCI calculations involved in each of the two MBE-FCI expansions in Figure \ref{h2o_dz_problem_2_fig} starting from a CASSCF reference, the expansion at $r = 4.0$ {\AA} involves only half as many as the corresponding expansion at $r = 2.6$ {\AA}, cf. Table \ref{h2o_dz_stretch_table}. This is indeed a general pattern in the case of H$_2$O, indicating that the dynamic correlation out of the CASSCF reference becomes confined to fewer orbitals as the bonds are stretched (as more contributions are screened at longer bond lengths). To underline this fact, Table \ref{h2o_dz_stretch_table} collects similar results for selected bond lengths all the way up to $r = 10.0$ {\AA}, employing the screening threshold of Figure \ref{h2o_dz_problem_2_fig} as well as one that is significantly tighter ($a=2.5$). As is clear from the numbers in Table \ref{h2o_dz_stretch_table}, FCI results to micro-Hartree accuracy are obtained even using an aggressive screening threshold ($a=5.0$), except for very long bond lengths ($r\geq6.0$ {\AA}), at which a tighter threshold is needed to converge to the FCI results with the same precision.
\begin{table}[ht]
\begin{center}
\begin{tabular}{r|rrrr|rr}
\toprule
\multicolumn{1}{c|}{\multirow{2}{*}{$r$ ({\AA})}} & \multicolumn{1}{c}{\multirow{2}{*}{$E_{\text{RHF}}$ ($E_{\text{H}}$)}} & \multicolumn{1}{c}{\multirow{2}{*}{$E_{\text{FCI}}$ ($E_{\text{H}}$)}} & \multicolumn{2}{c|}{$E_{\text{MBE-FCI}}$ ($E_{\text{H}}$)} & \multicolumn{2}{c}{$N_{\text{MBE-FCI}}$} \\
& & & \multicolumn{1}{c}{$a=5.0$} & \multicolumn{1}{c|}{$a=2.5$} & \multicolumn{1}{c}{$a=5.0$} & \multicolumn{1}{c}{$a=2.5$} \\
\midrule\midrule
$1.0$ & $-76.021444$ & $-76.239715$ & $-76.239717$ & $-76.239716$ & $18608$ & $34276$ \\
$1.4$ & $-75.838430$ & $-76.098115$ & $-76.098116$ & $-76.098116$ & $17626$ & $32810$ \\
$1.8$ & $-75.648669$ & $-75.976977$ & $-75.976978$ & $-75.976977$ & $16148$ & $30886$ \\
$2.2$ & $-75.510288$ & $-75.926497$ & $-75.926497$ & $-75.926497$ & $14530$ & $27633$ \\
$2.6$ & $-75.405741$ & $-75.913158$ & $-75.913157$ & $-75.913158$ & $12279$ & $22068$ \\
$3.0$ & $-75.333714$ & $-75.909947$ & $-75.909948$ & $-75.909946$ & $9343$ & $15910$ \\
$4.0$ & $-75.235531$ & $-75.908708$ & $-75.908708$ & $-75.908709$ & $6661$ & $15807$ \\
$6.0$ & $-75.175422$ & $-75.908627$ & $-75.908635$ & $-75.908627$ & $4312$ & $6785$ \\
$8.0$ & $-75.152537$ & $-75.908626$ & $-75.908618$ & $-75.908628$ & $3713$ & $4810$ \\
$10.0$ & $-75.138787$ & $-75.908626$ & $-75.908618$ & $-75.908626$ & $2606$ & $3341$ \\
\bottomrule
\end{tabular}
\end{center}
\caption{RHF, FCI, and MBE-FCI total energies for H$_2$O using a cc-pVDZ basis set at various O--H bond lengths ($r$). The MBE-FCI expansions are all based on CASSCF($8,6$) starting points and $N_{\text{MBE-FCI}}$ gives the corresponding number of individual CASCI calculations.}
\label{h2o_dz_stretch_table}
\end{table}

\subsubsection{Nitrogen Dimer Stretch}\label{n2_subsubsection}

\begin{figure}[ht]
\begin{center}
\includegraphics{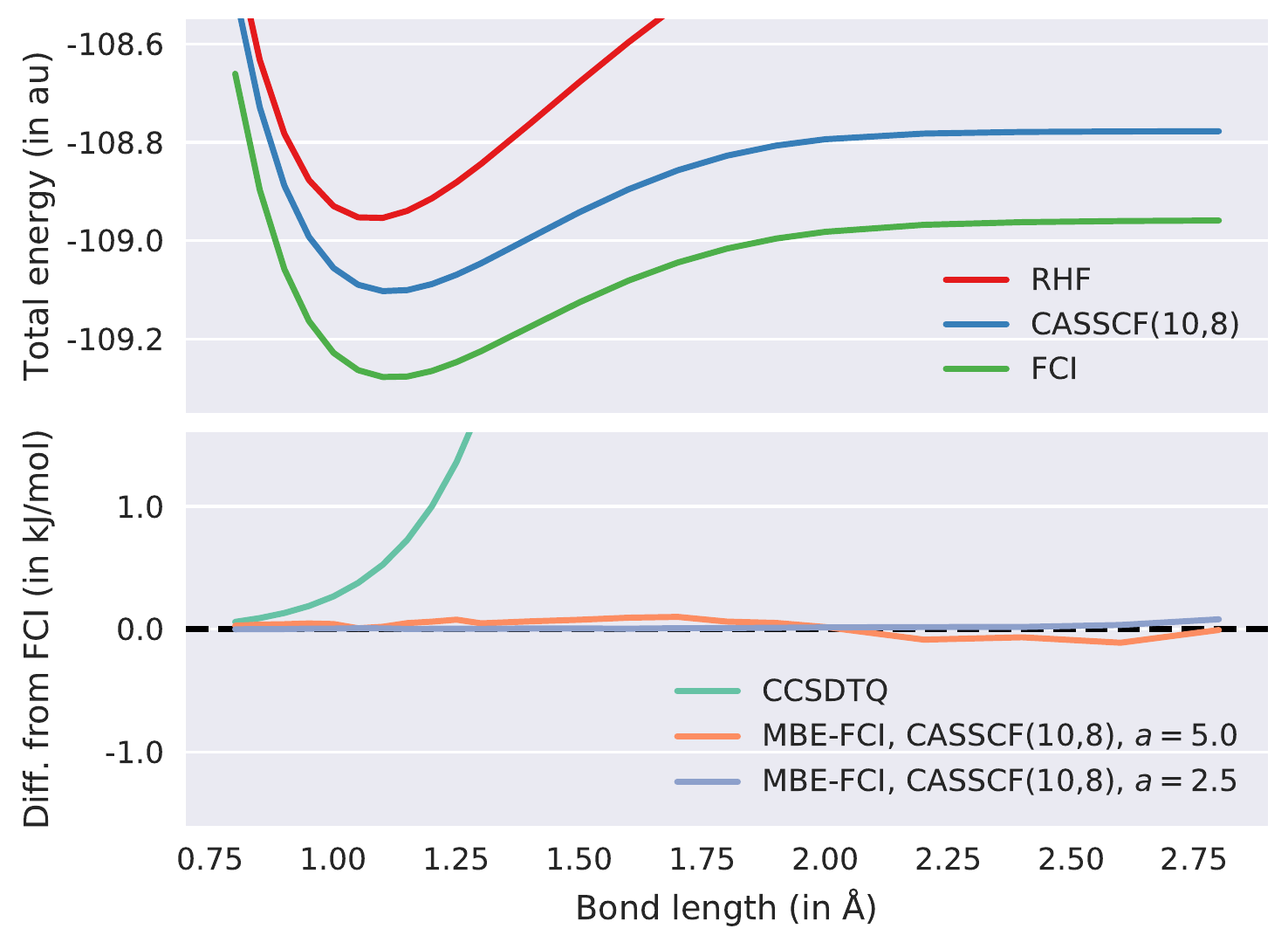}
\caption{Upper panel: RHF, CASSCF($10,8$), and FCI results for the $X$~${^{1}}\Sigma^{+}_g$ ground state of N$_2$ in a cc-pVDZ basis set. Lower panel: Deviation of the CCSDTQ and MBE-FCI results from FCI.}
\label{n2_dz_res_fig}
\end{center}
\end{figure}

Due to its multiple-bond dissociation, the nitrogen dimer has long served as a favoured stress test example for new multireference methods~\cite{olsen_bond_break_n2_jcp_2000,olsen_bond_break_n2_cpl_2001,chan_bond_break_n2_jcp_2004,hanrath_bond_break_n2_mol_phys_2009,legeza_bond_break_n2_pra_2018}, in part due to its manageable valence region of 10 electrons and in part due to its exceptionally strong bond at equilibrium geometry~\cite{huber_herzberg_geo_book}. In comparison with the symmetric stretch of the two single bonds in H$_2$O (Section \ref{h2o_stretch_subsubsection}), the alteration of the electronic structure of N$_2$ upon stretching its triple bond away from the weakly correlated, single-reference regime in the vicinity of the equilibrium structure takes place at a significantly more rapid pace, cf. the PEC in Figure \ref{n2_dz_res_fig}. Plotting once again the deviation with respect to the FCI reference data in the lower panel of Figure \ref{n2_dz_res_fig}, this transition between the two regimes is perhaps most easily identifiable from the rate at which the CCSDTQ solution diverges. As for H$_2$O, a simple valence space CASSCF solution captures the dissociation process in a qualitatively correct manner.\\

In assessing how the MBE-FCI method performs for this more challenging bond dissociation, Figure \ref{n2_dz_res_fig} further presents the deviation from FCI for two expansions which are both formulated on top of the CASSCF($10,8$) reference, but differ in the employed screening threshold. In comparison with the H$_2$O example in Figure \ref{h2o_dz_res_fig}, the $a = 5.0$ results are observed only to be slightly more in error. In terms of the mean absolute deviation (MAD) from FCI, this is $0.041$ kJ/mol, in comparison with a corresponding value of $0.005$ kJ/mol in the case of H$_2$O (different number of data points). The results obtained with the more conservative threshold of $a=2.5$, however, are quantitatively correct, thus supporting the conclusions drawn from the H$_2$O example in Table \ref{h2o_dz_stretch_table}. For these results, the MAD value reduces to $0.008$ kJ/mol.\\

\begin{table}[ht]
\begin{center}
\begin{tabular}{r|rrrr|rrrr}
\toprule
\multicolumn{1}{c|}{\multirow{3}{*}{order}} & \multicolumn{4}{c|}{$r = 1.10$ {\AA}} & \multicolumn{4}{c}{$r = 2.80$ {\AA}} \\
& \multicolumn{2}{c}{$\Delta E_{\text{FCI}}$ (kJ/mol)} & \multicolumn{2}{c|}{$N_{\text{MBE-FCI}}$} & \multicolumn{2}{c}{$\Delta E_{\text{FCI}}$ (kJ/mol)} & \multicolumn{2}{c}{$N_{\text{MBE-FCI}}$} \\
& \multicolumn{1}{c}{standard} & \multicolumn{1}{c}{$\pi$-pruned} & \multicolumn{1}{c}{standard} & \multicolumn{1}{c|}{$\pi$-pruned} & \multicolumn{1}{c}{standard} & \multicolumn{1}{c}{$\pi$-pruned} & \multicolumn{1}{c}{standard} & \multicolumn{1}{c}{$\pi$-pruned} \\
\midrule\midrule
CASSCF & $458.59$ & $458.59$ & $1$ & $1$ & $476.33$ & $476.33$ & $1$ & $1$ \\
\hline
$1$ & $183.00$ & $361.31$ & $18$ & $6$ & $215.49$ & $396.38$ & $18$ & $6$ \\
$2$ & $-14.04$ & $121.92$ & $153$ & $21$ & $-3.85$ & $156.90$ & $153$ & $21$ \\
$3$ & $-2.82$ & $45.55$ & $816$ & $56$ & $-11.60$ & $61.28$ & $816$ & $56$ \\
$4$ & $0.47$ & $-3.85$ & $3060$ & $120$ & $3.81$ & $-4.87$ & $3060$ & $120$ \\
$5$ & $0.01$ & $-3.20$ & $8358$ & $216$ & $-0.55$ & $-5.62$ & $8541$ & $216$ \\
$6$ & $0.02$ & $-0.62$ & $2564$ & $326$ & $-0.26$ & $0.20$ & $7175$ & $336$ \\
$7$ & & $-0.45$ & & $244$ & $-0.13$ & $-0.26$ & $116$ & $368$ \\
$8$ & & $0.13$ & & $32$ & & $0.15$ & & $80$ \\
$9$ & & $0.13$ & & $3$ & & $0.04$ & & $7$ \\
$10$ & & $0.12$ & & $2$ & & $0.02$ & & $5$ \\
\hline
Total & $0.02$ & $0.12$ & $14969$ & $1027$ & $-0.13$ & $0.02$ & $19879$ & $1216$ \\
\bottomrule
\end{tabular}
\end{center}
\caption{Deviation of MBE-FCI expansions ($a = 5.0$) from the FCI/cc-pVDZ results of Figure \ref{n2_dz_res_fig}. All expansions are based on a CASSCF($10,8$) reference, but differ in whether or not $\pi$-pruning has been enabled. As in Table \ref{h2o_dz_stretch_table}, $N_{\text{MBE-FCI}}$ gives the corresponding number of individual CASCI calculations at a given order in MBE-FCI.}
\label{n2_dz_table}
\end{table}
As mentioned earlier, additional measures have been implemented for the diatomics of the present study in order to warrant convergence onto a state of the proper symmetry (${^{1}}\Sigma^{+}_g$ in the present case). To illustrate the effect of the proposed $\pi$-pruning on MBE-FCI calculations, Table \ref{n2_dz_table} reports results for two selected N--N bond lengths, one near equilibrium ($r = 1.1$ {\AA}) and one toward the dissociated limit ($r = 2.8$ {\AA}). Generally, three conclusions may be drawn from the examples in Table \ref{n2_dz_table}. First, the use of $\pi$-pruning results in a significantly reduced total number of involved CASCI calculations, and second, this prescreening of contributions to the MBE happens at the expense of convergence being met at slightly higher orders in the expansion and at an overall slightly slower rate. Third, the accuracy is in general not compromised by pruning and in many instances the residual error against FCI is even reduced (as for the strongly correlated example at $r = 2.8$ {\AA}, cf. also the C$_2$ case in Section \ref{c2_subsubsection}). All three features may be attributed to the manner in which degenerate $\pi$-orbitals are always included in pairs when pruning is enabled. Thus, excitations into any given pair of $\pi$-orbitals outside the reference space are first accounted for at order $2$, cf. Table \ref{n2_dz_table}, and excitations into mixed $\pi_u$/$\pi_g$ orbitals will not occur until order $4$. However, the states, onto which the corresponding CASCI calculations convergence, are guaranteed to be of the proper symmetry. Ultimately, in the case of N$_2$, the use of $\pi$-pruning hence works to compress the MBE-FCI even further with a corresponding reduction in the involved time-to-solution.

\subsubsection{Carbon Dimer Stretch}\label{c2_subsubsection}

\begin{figure}[htbp]
\begin{center}
\includegraphics{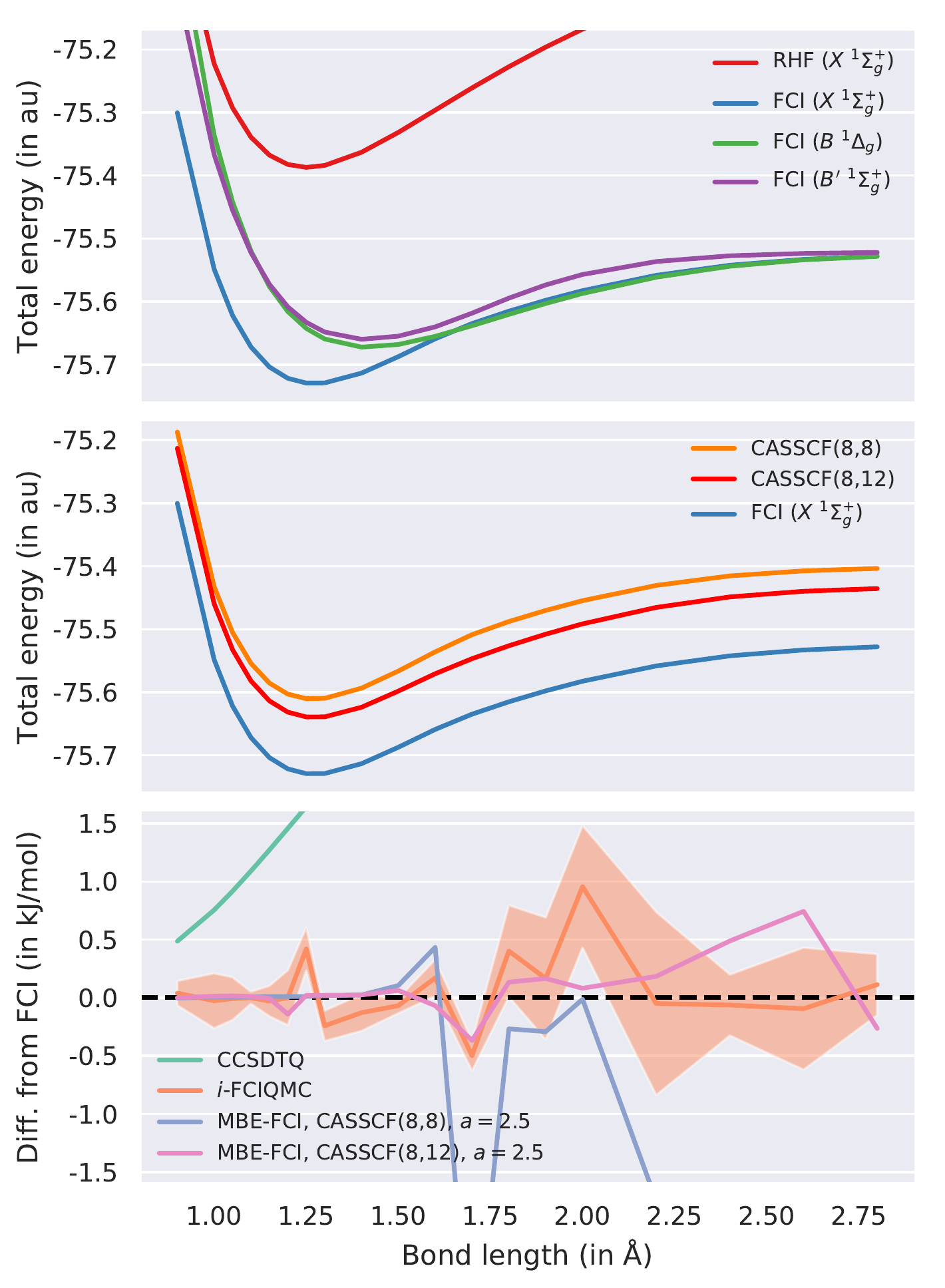}
\caption{Upper panel: RHF and FCI results for the $X$~${^{1}}\Sigma^{+}_g$, $B$~${^{1}}\Delta_g$, and $B^{\prime}$~${^{1}}\Sigma^{+}_g$ states of C$_2$ in a cc-pVDZ basis set. Middle panel: State-averaged CASSCF results for the $X$~${^{1}}\Sigma^{+}_g$ state. Lower panel: Deviation of the CCSDTQ, $i$-FCIQMC~\cite{booth_alavi_fciqmc_bond_break_c2_jcp_2011}, and MBE-FCI results from FCI ($X$~${^{1}}\Sigma^{+}_g$).}
\label{c2_dz_res_fig}
\end{center}
\end{figure}

The interpretation and classification of the electronic structure and molecular bond pattern in the carbon dimer easily rank among the most disputed in the literature for a simple diatomic molecule~\cite{shaik_hiberty_c2_dispute_nature_chem_2012,grunenberg_c2_dispute_nature_chem_2012,shaik_hoffmann_c2_dispute_angew_chem_2013,frenking_hermann_c2_dispute_angew_chem_2013}. Even at equilibrium geometry, its electronic structure shows significant multireference character~\cite{abrams_sherrill_bond_break_c2_jcp_2004,sherrill_piecuch_bond_break_c2_jcp_2005,varandas_bond_break_c2_jcp_2008,booth_alavi_fciqmc_bond_break_c2_jcp_2011,sharma_dmrg_bond_break_c2_jcp_2015,holmes_umrigar_sharma_shci_bond_break_c2_jcp_2017}, and as may be recognized from the PECs of the three lowest-lying singlet states in the upper panel of Figure \ref{c2_dz_res_fig}, any attempt to describe the stretch of C$_2$ will have to overcome an allowed crossing between a ${^{1}}\Sigma^{+}_g$ and a ${^{1}}\Delta_g$ state immediately followed by an avoided crossing of the two lowest states of ${^{1}}\Sigma^{+}_g$ symmetry. All three states rapidly approach the same asymptotic limit, namely the product of two C(${^{3}}P$) states.\\

Contrary to Sections \ref{h2o_stretch_subsubsection} and \ref{n2_subsubsection}, we will here report MBE-FCI results using more than a single choice of multideterminantal expansion reference. Namely, two different references have been employed, each of which have been constructed on the basis of point-group arguments. The CASSCF($8,8$) reference includes all valence orbitals that emerge from the 2s and 2p AOs (2 sets of bonding and antibonding $\sigma$ orbitals and degenerate sets of $\pi_{u}$/$\pi_{g}$ orbitals), while the CASSCF($8,12$) reference adds an additional set of $\pi_{u}$ and $\pi_{g}$ orbitals. Due to the closeness of the states in Figure \ref{c2_dz_res_fig}, both CASSCF references have been obtained as state-averaged solutions over three $A_{g}$ states and a single $B_{1g}$ state in the $D_{2\text{h}}$ subgroup. The middle panel of Figure \ref{c2_dz_res_fig} compares these CASSCF solutions.\\

Both sets of MBE-FCI results, alongside the $i$-FCIQMC results from Ref. \citenum{booth_alavi_fciqmc_bond_break_c2_jcp_2011} and the CCSDTQ results of the present work, are presented in the lower panel of Figure \ref{c2_dz_res_fig}. Starting with the results for the MBE-FCI expansion formulated on top of the valence space CASSCF($8,8$) reference, these differ in their quality in the intervals before and after the onset of the (avoided) state crossings. In the former ($r < 1.6$ {\AA}), the excellent performance for the N$_2$ case in Section \ref{c2_subsubsection} is repeated, while in the latter ($r \geq 1.6$ {\AA}), a clear deterioration of the results is observed with deviations of $\delta E = -1.81$ kJ/mol and $\delta E = -3.96$ kJ/mol at $r = 2.0$ {\AA} and $r = 2.2$ {\AA}, respectively. For the other MBE-FCI expansion, thermochemical agreement with the FCI reference is maintained throughout the range of tested bond lengths with an MAD value of $0.152$ kJ/mol, although also these results are generally more in error at longer bond lengths. In comparison with the $i$-FCIQMC results, the latter set of MBE-FCI results is hence comparable to, if not an improvement upon these for all tested bond lengths. In that respect, we note that even $i$-FCIQMC results were hard to obtain at longer distances. As noted in Ref. \citenum{booth_alavi_fciqmc_bond_break_c2_jcp_2011}, at bond lengths of $r \geq 2.2$ {\AA} the energies were determined from the averaged value of the so-called shift parameter due to the absence of a significantly weighted determinant to project onto whereas the projected energy was used $r < 2.2$ {\AA}, using the largest weighted spin-coupled function as a reference (not necessarily the HF determinant).\\

\begin{figure}[ht]
\begin{center}
\includegraphics{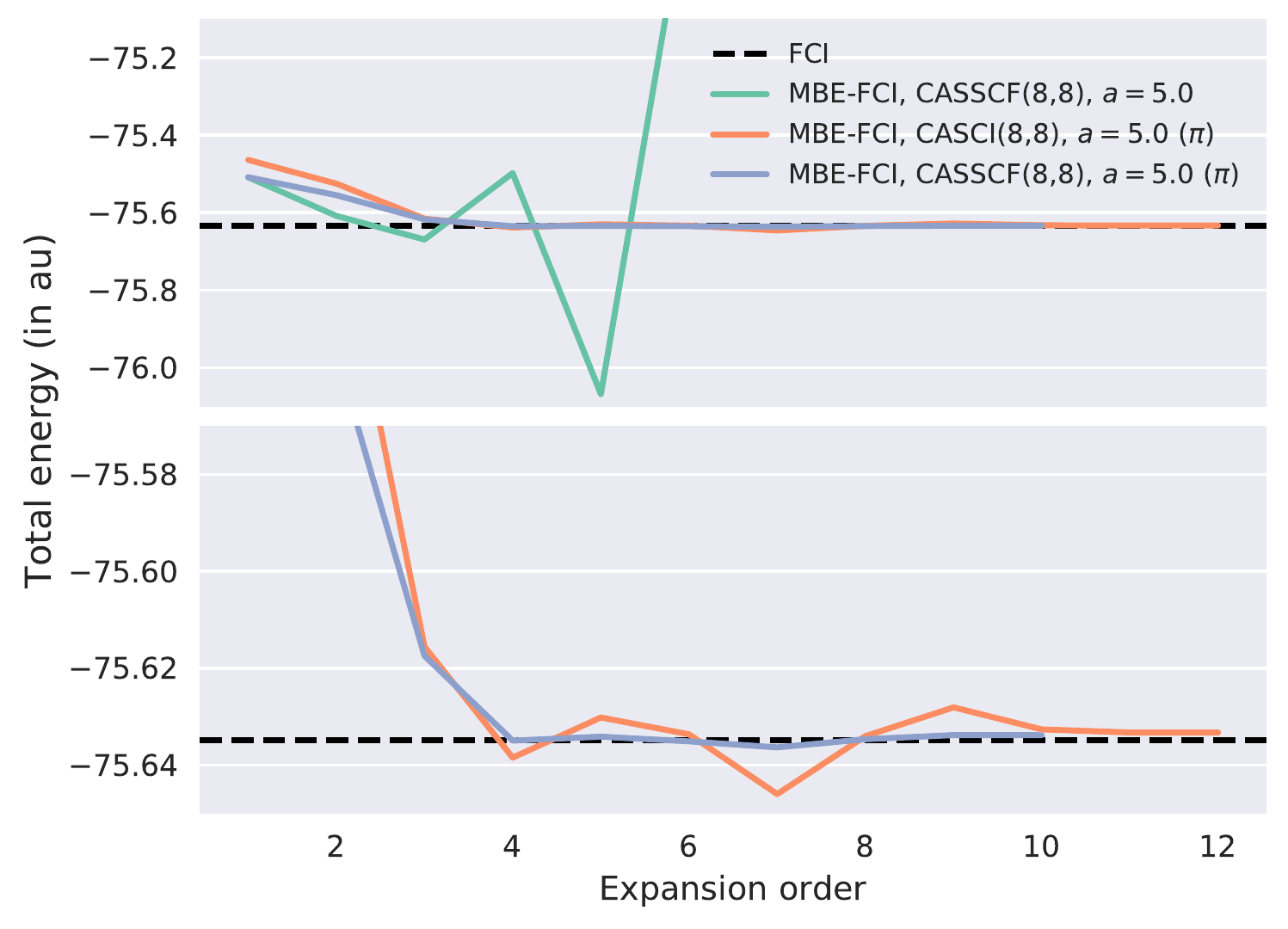}
\caption{Energy convergence of MBE-FCI expansions ($a = 5.0$) with and without $\pi$-pruning for the $X$~${^{1}}\Sigma^{+}_g$ state of C$_2$ at a bond distance of $r = 1.7$ {\AA}. The lower panel presents the results of the expansions with $\pi$-pruning enabled in a focused interval around the FCI reference result.}
\label{c2_dz_prune_fig}
\end{center}
\end{figure}
Given the results in Figure \ref{c2_dz_res_fig}, it is clear that the question of how to choose an optimal reference space for MBE-FCI in the case of C$_2$ remains unresolved, and the results give further evidence to the fact that what constitutes an ideal reference space within MBE-FCI need not necessarily coincide with an ideal multideterminantal basis in traditional multireference theory. As mentioned at the end of Section \ref{theory_section}, the optimal, preferable black-box selection of MBE-FCI reference spaces still warrants more investigation. However, even the results of the MBE-FCI expansion formulated on top of the simple CASSCF(8,8) reference would be essentially indistinguishable from the corresponding FCI results if these were both plotted on the scale used for the PECs in the upper two panels of Figure \ref{c2_dz_res_fig}. To emphasize the merit of even this simple reference space, Figure \ref{c2_dz_prune_fig} reports the convergence of three different MBE-FCI expansions at the point of the (avoided) state crossings. The expansions all use a looser screening threshold ($a = 5.0$) than that used in Figure \ref{c2_dz_res_fig}, while differing in what orbitals are used (HF or CASSCF) and whether or not $\pi$-pruning has been used. Two conclusions may be drawn from the convergence profiles in Figure \ref{c2_dz_res_fig}. First, given the closeness of the three states in questions, which are all spanned by the same irreducible representation ($A_g$) in the $D_{2\text{h}}$ point-group, the use of $\pi$-pruning is not a mere convenience, as for N$_2$ in Section \ref{n2_subsubsection}, but rather of crucial importance in converging onto the requested $X$~${^{1}}\Sigma^{+}_g$ state. Second, it is not the employed MOs, but rather the employed reference space which results in proper convergence. The use of optimized orbitals, however, such as those of a preceding CASSCF calculation, is observed to accelerate convergence, as was previously noted in connection with the use of CC natural orbitals for weakly correlated systems in Ref. \citenum{eriksen_mbe_fci_weak_corr_jctc_2018}.

\subsubsection{Hydrogen Chain Dissociation}\label{h10_subsubsection}

\begin{figure}[ht]
\begin{center}
\includegraphics{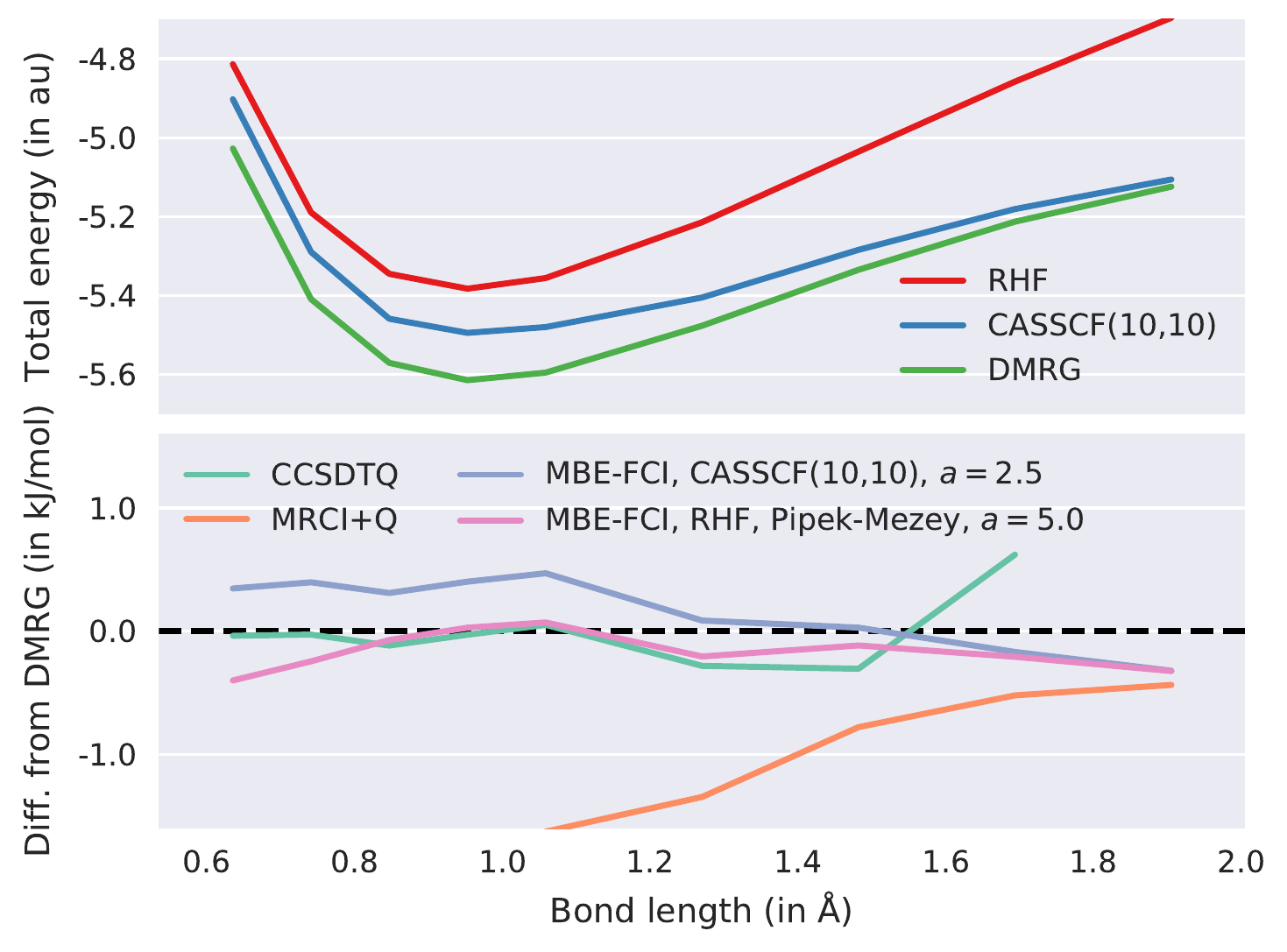}
\caption{Upper panel: RHF, CASSCF, and DMRG results for the ${^{1}}\Sigma^{+}_g$ ground state of a chain of 10 hydrogen atoms in a cc-pVDZ basis set. Lower panel: Deviation of the MRCI+Q, CCSDTQ, and MBE-FCI results from DMRG.}
\label{h10_chain_dz_res_fig}
\end{center}
\end{figure}

Recently, hydrogen model systems such as chains, rings, and sheets have gathered attention as prototypical systems for evaluating and calibrating various quantum chemical schemes in their description of concurrent bond breaking processes~\cite{chan_drmg_bond_break_hydrogen_jcp_2006,zhang_krakauer_afqmc_bond_break_hydrogen_jcp_2007,scuseria_hf_bond_break_hydrogen_jcp_2009,mazziotti_2rdm_bond_break_hydrogen_jcp_2010,reichman_dmft_bond_break_hydrogen_prl_2011,rubio_vmc_bond_break_hydrogen_prb_2011,simons_collab_bond_break_hydrogen_prx_2017}. Among the chain systems, the simultaneous stretching of all bonds of the tenfold linear chain and the associated PEC in the upper panel of Figure \ref{h10_chain_dz_res_fig} has emerged as an affordable minimal test case which contains all of the physics that occurs in the thermodynamic limit, i.e., the case of a chain built from an infinite number of H atoms. For the present study, however, it will suffice to note that the usual transition from weak to strong correlation discussed in the previous sections again takes place, although on a significantly more extended scale.\\

In a recent benchmark study~\cite{simons_collab_bond_break_hydrogen_prx_2017}, errors of the standard MRCI+Q method with respect to corresponding FCI results were found to be more or less uniform along the entire PEC for the H$_{10}$ chain, i.e., in the micro-Hartree range. Although these results were obtained in a minimal STO-6G basis, similar conclusions were drawn for corresponding results obtained with a larger cc-pVDZ basis set upon comparing the MRCI+Q results with those from DMRG calculations. For that reason and its applicability in larger basis sets, MRCI+Q was chosen as the reference for the study of linear H$_{10}$ in Ref. \citenum{simons_collab_bond_break_hydrogen_prx_2017} with basis sets of up to pentuple-$\zeta$ quality. However, by plotting the MRCI+Q/cc-pVDZ results against DMRG, cf. the lower panel of Figure \ref{h10_chain_dz_res_fig}, and by further comparing them to corresponding CCSDTQ results, one realizes that the errors of MRCI+Q are in fact non-uniform and non-negligible, in particular in the weakly correlated regime in the vicinity of the equilibrium H--H distance of $r \sim 0.95$ {{\AA}. This is not entirely unexpected as the treatment of dynamic correlation out of the ($10,10$) active space in the MRCI+Q method remains somewhat limited. On the other hand, near the largest of the tested bond distances in Ref. \citenum{simons_collab_bond_break_hydrogen_prx_2017}, the correlation is primarily static and the MRCI+Q results are likely to provide an adequate reference. This observation is further supported by the fact that the results of CASSCF($10,10$) calculations, on top of which the MRCI+Q calculations are performed, asymptotically converge to the physically correct dissociation limit (see upper panel of Figure \ref{h10_chain_dz_res_fig}).\\

Figure \ref{h10_chain_dz_res_fig} further presents MBE-FCI results obtained starting from the same CASSCF($10,10$) reference. These results are seen to lie slightly above the variational DMRG results at short bond lengths, while the are likely to mark an improvement over the DMRG values by a similar amount at larger distances as they are seen to agree with the MRCI+Q results in the limit of strong correlation. However, despite its promising performance for the present H$_{10}$ case, the MBE-FCI method will hold limited promise of providing detailed information in the thermodynamic limit ($N\rightarrow\infty$) of the system whenever it has to rely on CASSCF($N,N$) expansion references. Furthermore, the canonical orbitals of such calculations will remain delocalized over large sections of the chain, a fact which in turn inhibits the orbital screening used in the MBE-FCI method. To that end, it is worth noting that the DMRG treatment employs a split-localized basis consisting of separately localized occupied and virtual orbitals~\cite{chan_dmrg_2015}, which significantly facilitates the description of dissociated systems such as the present case with individual hydrogen atom entities as the end products.\\

\begin{figure}[ht]
\begin{center}
\includegraphics{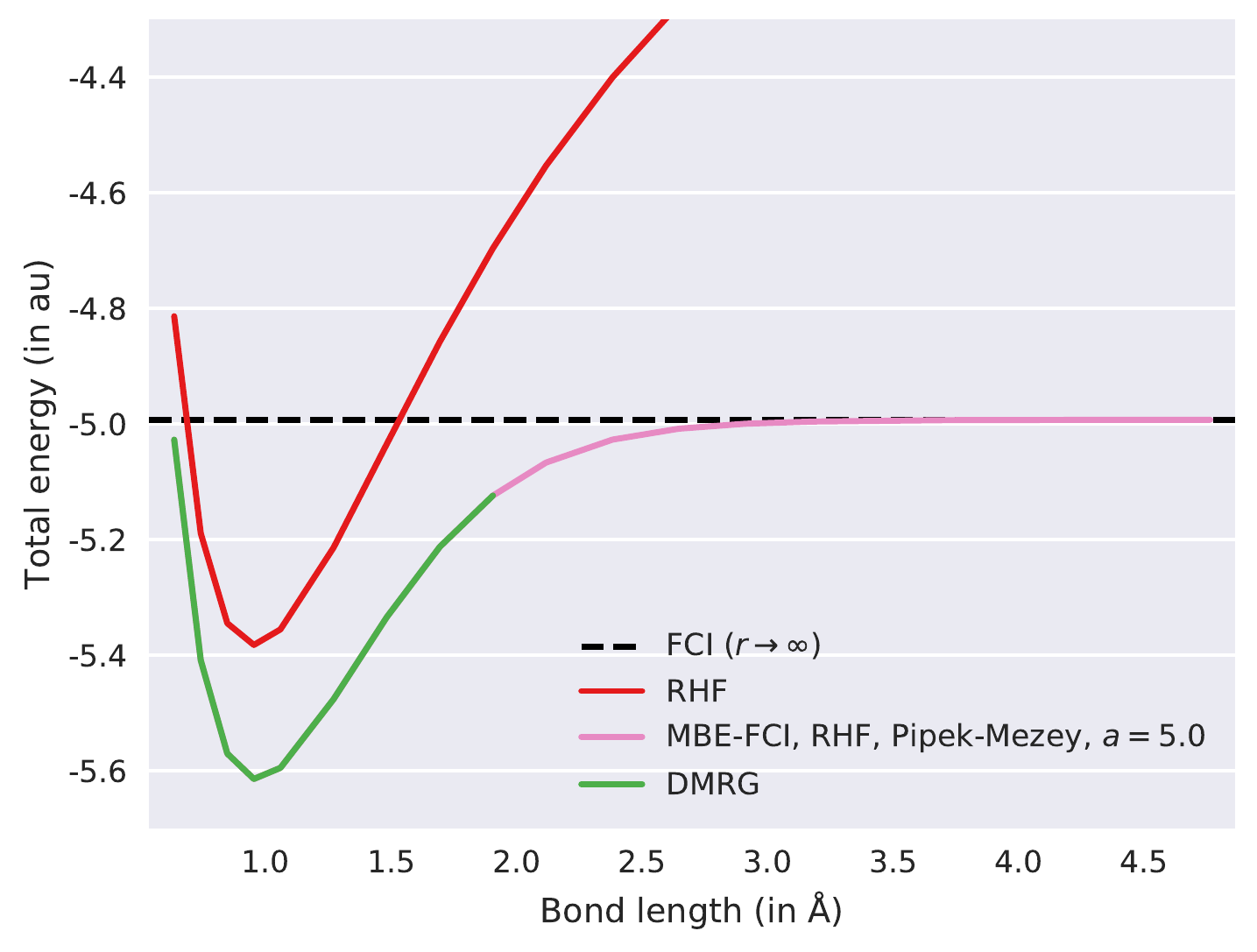}
\caption{The PEC for the ${^{1}}\Sigma^{+}_g$ ground state of a H$_{10}$ chain in a cc-pVDZ basis set, as calculated by RHF, DMRG, and MBE-FCI using an RHF expansion reference and Pipek-Mezey localized virtual MOs. The dashed line marks the FCI complete dissociation limit, $E=-4.992784$ $E_{\text{H}}$.}
\label{h10_chain_dz_pm_fig}
\end{center}
\end{figure}
Pursuing a similar strategy, the lower panel of Figure \ref{h10_chain_dz_res_fig} also includes results for an MBE-FCI expansion starting from an RHF determinant and using localized Pipek-Mezey rather than canonical virtual orbitals~\cite{pipek_mezey_jcp_1989}. Despite the difficulties associated with confining the spatial extent of virtual orbitals~\cite{ida_loc_orb_review_tca_2014}, the use of localized virtual MOs is seen from Figure \ref{h10_chain_dz_res_fig} to give excellent results for all but the shortest bond distances in the repulsive region where the concept of locality is anyways somewhat ill-defined. In particular, the MBE-FCI results using localized orbitals are observed to converge toward the correct limit as the bonds are stretched. It should be noted that these results are obtained using a looser threshold ($a = 5.0$) than the corresponding MBE-FCI results obtained by starting from the CASSCF(10,10) reference and the calculations are run in $C_1$ rather than $D_{2\text{h}}$ symmetry. In fact, the MBE-FCI expansions using localized orbitals converge at an increasingly rapid pace as the 9 involved bonds are stretched, analogous to the symmetric stretching of H$_2$O in Section \ref{h2o_stretch_subsubsection}, making it trivial to extend the range of tested bond lengths, cf. Figure \ref{h10_chain_dz_pm_fig}. Furthermore, we note that because of its standard formulation in terms of an RHF expansion reference, the performance of MBE-FCI in Figure \ref{h10_chain_dz_pm_fig} will be transferrable to larger chains and basis sets, thus offering a viable approach for the treatment of the thermodynamic limit. However, such investigations fall outside the scope of the present work and are hence postponed to future studies.

\subsection{Large-Scale Calculations}\label{h2o_c2_equil_subsection}

Having assessed the use of multideterminantal expansion references for strongly correlated systems using modest basis sets of double-$\zeta$ quality, we will now discuss how these may further aid in the application of the MBE-FCI method when using extended basis sets. Unlike in Section \ref{strong_corr_subsection}, which illustrated the performance of the method for structurally distorted molecules, the following two examples in Sections \ref{h2o_equil_subsubsection} (H$_2$O) and \ref{c2_equil_subsubsection} (C$_2$) will be concerned with molecules at their equilibrium geometry in order to facilitate direct comparisons with alternative methods. We stress, however, that the convergence profiles and trends of the MBE-FCI method have previously been shown to be practically independent of the size of the basis set employed~\cite{eriksen_mbe_fci_jpcl_2017,eriksen_mbe_fci_weak_corr_jctc_2018}. That is, given that the method converges in a double-$\zeta$ basis set, the same is bound to hold true in basis sets of larger cardinal numbers as well.\\

In the following, we will once again compare the performance of MBE-FCI to CCSDTQ, while for C$_2$ in Section \ref{c2_equil_subsubsection}, we will furthermore relate our results to those of corresponding $i$-FCIQMC calculations and results obtained by the method of composite correlation energy extrapolation
through intrinsic scaling FCI (CEEIS-FCI)~\cite{bytautas_ruedenberg_ceeis_jcp_2004}.

\subsubsection{Water at Equilibrium Geometry}\label{h2o_equil_subsubsection}

\begin{table}[ht]
\begin{center}
\begin{tabular}{l|cccc|cc}
\toprule
\multicolumn{1}{c|}{Method} & \multicolumn{1}{c}{cc-pVDZ} & \multicolumn{1}{c}{cc-pVTZ} & \multicolumn{1}{c}{cc-pVQZ} & \multicolumn{1}{c|}{cc-pV5Z} & \multicolumn{1}{c}{CBS(TZ/QZ)} & \multicolumn{1}{c}{CBS(QZ/5Z)} \\
\midrule\midrule
CCSD(T) & $-214.06$ & $-275.33$ & $-295.44$ & $-302.58$ & $-310.11$ & $-310.07$ \\
CCSDTQ & $-214.79$ & $-275.36$ & $-295.29$ & $-302.24$ & $-309.83$ & $-309.53$ \\
\hline
MBE-FCI-{\textbf{a}} & $-214.80$ & $-275.40$ & $-295.15$ & \multicolumn{1}{c|}{N/A} & $-309.56$ & \multicolumn{1}{c}{N/A} \\
MBE-FCI-{\textbf{b}} & $-214.80$ & $-275.35$ & $-295.28$ & $-302.27$ & $-309.82$ & $-309.60$ \\
\bottomrule
\end{tabular}
\end{center}
\caption{MBE-FCI correlation energies (in m$E_{\text{H}}$) using HF ({\textbf{a}}) or CASCI($8,6$) ({\textbf{b}}) expansion references for the ${^{1}}A_{1}$ ground state of H$_2$O when using cc-pV$X$Z ($X$ = D, T, Q, and 5) basis sets in comparison with CCSD(T) and CCSDTQ results. The screening threshold used is $a = 5.0$ and a CCSD(T) base is used. The FCI/cc-pVDZ reference result is $-214.80$ m$E_{\text{H}}$.}
\label{h2o_equil_table}
\end{table}
As already mentioned at the end of Section \ref{intro_section}, the use of multideterminantal expansion references in MBE-FCI is not restricted to the treatment of strong correlation, as they might also be considered as a way to provide focussed compressions of the involved orbital expansions in the case of weak correlation. The reason for this is simply that a large number of possible contributions to these expansions get excluded from the decompositions of the FCI correlation energy compared to when the reference space is comprised of only a single determinant (and the expansion space is correspondingly larger). To that end, Table \ref{h2o_equil_table} presents MBE-FCI results, all using a CC with perturbative triples (CCSD(T)) expansion base model~\cite{original_ccsdpt_paper,ccsdpt_perturbation_stanton_cpl_1997}, for H$_2$O (at the equilibrium geometry used in Ref. \citenum{eriksen_mbe_fci_weak_corr_jctc_2018}) with basis sets ranging from double- to pentuple-$\zeta$ quality, corresponding to $23$--$200$ MOs. For reference and clarity, the underlying CCSD(T) results are also presented in Table \ref{h2o_equil_table}. Whereas the smaller basis sets correspond to modest-sized MBE-FCI expansions, the larger basis sets, in particular cc-pV5Z, give rise to calculations that are well beyond the application range of the MBE-FCI method in its original formulation.\\

Despite the fundamental difference in how the CCSDTQ method approximates and the MBE-FCI method approaches FCI (the former is defined by means of a truncated cluster ansatz whereas the latter sacrifices accuracy through its screening algorithm), it is instructive to compare the results of the two methods for the present case. As is recognized from Table \ref{h2o_equil_table}, the CCSDTQ results differ only marginally from those obtained with MBE-FCI using either an HF (MBE-FCI-{\textbf{a}})~\bibnote{The H$_2$O/cc-pV5Z calculation using MBE-FCI starting from an HF expansion reference was not performed due to time limitations.} or a CASCI($8,6$) (MBE-FCI-{\textbf{b}}) expansion reference. As discussed in Ref. \citenum{eriksen_mbe_fci_weak_corr_jctc_2018}, CCSDTQ calculations for a single-reference system like H$_2$O typically yield results that are in close agreement with FCI (for H$_2$O, expected errors of $\sim 10^{-5}$ $E_{\text{H}}$). Hence, particularly for the MBE-FCI-{\textbf{b}} expansions, with respect to which the CCSDTQ results never disagree by more than $0.03$ m$E_{\text{H}}$, the non-variational nature of both methods and the relative loose MBE-FCI threshold make it is next to impossible to determine which of the two sets of results lie closest to the exact answer.\\

In Table \ref{h2o_equil_table}, we have further extrapolated the correlation energies to the complete basis set (CBS) limit using the two-point formula of Helgaker {\it{et al.}}~\cite{helgaker_basis_conv_jcp_1997} and the results of two successive basis sets. As expected, the CBS(QZ/5Z) results for the CCSDTQ and MBE-FCI-{\textbf{b}}  methods agree very well (difference of $0.07$ m$E_{\text{H}}$), while the CCSD(T) method overestimates the correlation energy by approximately $-0.5$ m$E_{\text{H}}$ in this limit, that is, a difference well in excess of $1$ kJ/mol. In this context, it should be noted that the magnitude of the (T) triples correction---when compared to FCI---is considerably different for basis sets of varying sizes. For example, the differences of CCSD(T) against MBE-FCI-{\textbf{b}} range from $+0.74$ m$E_{\text{H}}$ in the cc-pVDZ basis set to $-0.31$ m$E_{\text{H}}$ in the larger cc-pV5Z basis set, which renders the extrapolation to the CBS limit of somewhat limited value. On the other hand, the significantly more advanced CCSDTQ method remains stable against MBE-FCI upon moving to larger basis sets, thus yielding reliable results even in the extrapolated CBS limit.

\subsubsection{Carbon Dimer at Equilibrium Geometry}\label{c2_equil_subsubsection}

\begin{table}[ht]
\begin{center}
\begin{tabular}{l|llll|ll}
\toprule
\multicolumn{1}{c|}{Method} & \multicolumn{1}{c}{cc-pVDZ} & \multicolumn{1}{c}{cc-pVTZ} & \multicolumn{1}{c}{cc-pVQZ} & \multicolumn{1}{c|}{cc-pV5Z} & \multicolumn{1}{c}{CBS(TZ/QZ)} & \multicolumn{1}{c}{CBS(QZ/5Z)} \\
\midrule\midrule
CCSD(T) & $-339.79$ & $-381.62$ & $-395.04$ & $-399.59$ & $-404.83$ & $-404.36$ \\
CCSDTQ & $-341.03$ & $-382.97$ & $-396.26$ & $-400.68$ & $-405.96$ & $-405.32$ \\
\hline
$i$-FCIQMC & $-341.60(10)$ & $-383.55(10)$ & $-396.53(30)$ & \multicolumn{1}{c|}{N/A} & $-406.00(59)$ & \multicolumn{1}{c}{N/A} \\
CEEIS-FCI & $-341.65$ & $-383.52$ & $-397.03(30)$ & \multicolumn{1}{c|}{N/A} & $-406.89(52)$ & \multicolumn{1}{c}{N/A} \\
\hline
MBE-FCI-{\textbf{a}} & $-341.66$ & $-383.44$ & $-397.17$ & $-401.66$ & $-407.19$ & $-406.37$ \\
MBE-FCI-{\textbf{b}} & $-341.64$ & $-383.55$ & $-396.98$ & $-401.92$ & $-406.78$ & $-407.10$ \\
MBE-FCI-{\textbf{c}} & $-341.65$ & $-383.59$ & $-396.82$ & $-401.16$ & $-406.47$ & $-405.71$ \\
\bottomrule
\end{tabular}
\end{center}
\caption{MBE-FCI correlation energies (in m$E_{\text{H}}$) using CASCI($8,8$) ({\textbf{a}}) or CASSCF($8,8$) ({\textbf{b}} and {\textbf{c}}) expansion references for the $X$~${^{1}}\Sigma^{+}_g$ ground state of C$_2$ when using cc-pV$X$Z ($X$ = D, T, Q, and 5) basis sets in comparison with CCSD(T), CCSDTQ, $i$-FCIQMC~\cite{cleland_booth_alavi_fciqmc_jctc_2012}, and CEEIS-FCI~\cite{bytautas_ruedenberg_ceeis_jcp_2005} results. The screening thresholds used are $a = 5.0$ ({\textbf{a}} and {\textbf{b}}) or $a = 2.5$ ({\textbf{c}}) and a CCSD(T) base is used for MBE-FCI-{\textbf{a}}. The FCI/cc-pVDZ reference result is $-341.65$ m$E_{\text{H}}$, and uncertainties in the last digit of the $i$-FCIQMC and CEEIS-FCI results are given in parentheses where available.}
\label{c2_equil_table}
\end{table}
We now turn to the case of C$_2$, for which results are presented in Table \ref{c2_equil_table} obtained using the experimental equilibrium geometry of Douay {\it{et al.}} ($r_e = 1.24244$ {\AA})~\cite{douay_bernath_c2_exp_geo_1988}. Here, the basis set used in the correlated treatment ranges in size from $26$ MOs (cc-pVDZ) to $180$ MOs (cc-pV5Z) and the results are this time compared to those obtained with $i$-FCIQMC~\bibnote{The $i$-FCIQMC results of Table \ref{c2_equil_table} are those of Ref. \citenum{cleland_booth_alavi_fciqmc_jctc_2012}, not the original source in Ref. \citenum{booth_alavi_fciqmc_bond_break_c2_jcp_2011}, as the results of the latter were erroneously averaged and hence suffering from a systematic error. For more details, see the short comment on this in Ref. \citenum{cleland_booth_alavi_fciqmc_jctc_2012}.} and CEEIS-FCI~\cite{bytautas_ruedenberg_ceeis_jcp_2005}. As will be discussed below, the use of the CCSD(T) method as a base model (MBE-FCI-{\textbf{a}}) is somewhat less successful for the calculations on C$_2$ than what was observed earlier for H$_2$O (generally larger differences from the corresponding MBE-FCI-{\textbf{b}}/-{\textbf{c}} results). This observation may be rationalized by the fact that the CCSD(T) method itself is a significantly less reliable approximation to FCI in the case of C$_2$. In comparison with H$_2$O, the larger deviations of the CC results from MBE-FCI, as already discussed for CCSDTQ in Section \ref{c2_subsubsection}, is caused by the rather strong multireference character of the $X$~${^{1}}\Sigma^{+}_g$ ground state of C$_2$ even at equilibrium geometry. For instance, for the cc-pVDZ basis set, the differences from FCI amount to $+1.86$ m$E_{\text{H}}$ and $+0.62$ m$E_{\text{H}}$ for the CCSD(T) and CCSDTQ methods, respectively, which should be compared to the substantially smaller deviations observed for H$_2$O.\\

However, by switching from a CASCI (MBE-FCI-{\textbf{a}}) to a CASSCF (MBE-FCI-{\textbf{b}}/-{\textbf{c}}) expansion reference---in the absence of a base model but once again employing a simple ($8,8$) valence active space---the MBE-FCI method performs uniformly well across the basis sets that differ in size by almost an order of magnitude. Even with the loosest ($a = 5.0$) of the two tested screening thresholds in Table \ref{c2_equil_table} (MBE-FCI-{\textbf{b}}), the results match or improve upon those of the $i$-FCIQMC and CEEIS-FCI methods in the cc-pVTZ and cc-pVQZ basis sets. By tightening the threshold ($a = 2.5$) as in the MBE-FCI-{\textbf{c}} results, a further improvement is observed.\\

Upon extrapolating all of the results to the CBS limit and using the MBE-FCI-{\textbf{c}} results as a reference, a number of things should be noted: (i) the CBS extrapolations are generally more sensitive to the cardinal numbers entering the expressions than for H$_2$O as the CBS(TZ/QZ) and CBS(QZ/5Z) results show a larger variance; (ii) the CCSDTQ, $i$-FCIQMC, and CEEIS-FCI results at the CBS(TZ/QZ) limit all appear to be in error by roughly $0.5$ m$E_{\text{H}}$ ($1$--$2$ kJ/mol), whereas this error is significantly enlarged at the CCSD(T) level; and (iii) the CBS-extrapolated MBE-FCI-{\textbf{b}} results, albeit considerably cheaper to obtain, are also significantly in error in the largest cc-pV5Z basis set, illustrating the importance of a tight screening protocol for systems like C$_2$ with static correlation effects present in the ground state. In particular, the MBE-FCI-{\textbf{b}} result in the CBS(QZ/5Z) limit is lower than the corresponding result in the CBS(TZ/QZ) limit, whereas all other methods in Table \ref{c2_equil_table} show the opposite trend. The differences between the CBS-extrapolated CCSDTQ and MBE-FCI-{\textbf{c}} results, however, are roughly the same in both limits, serving as another indication of the quality of the latter set of results. This is further supported by comparing the results to a DMRG result~\cite{sharma_dmrg_bond_break_c2_jcp_2015} of $-396.92$ m$E_{\text{H}}$ in the cc-pVQZ basis set and semistochastic heat-bath CI results~\cite{holmes_umrigar_sharma_shci_bond_break_c2_jcp_2017} of $-396.95$ m$E_{\text{H}}$ and $-401.37$ m$E_{\text{H}}$ in the cc-pVQZ and cc-pV5Z basis sets, respectively. These results, however, were obtained at a slightly different equilibrium geometry of $r_e = 1.24253$ {\AA}.

\subsubsection{Parallel Scaling of the MBE-FCI Method}\label{parallel_scaling_subsubsection}

\begin{figure}[ht]
\begin{center}
\includegraphics{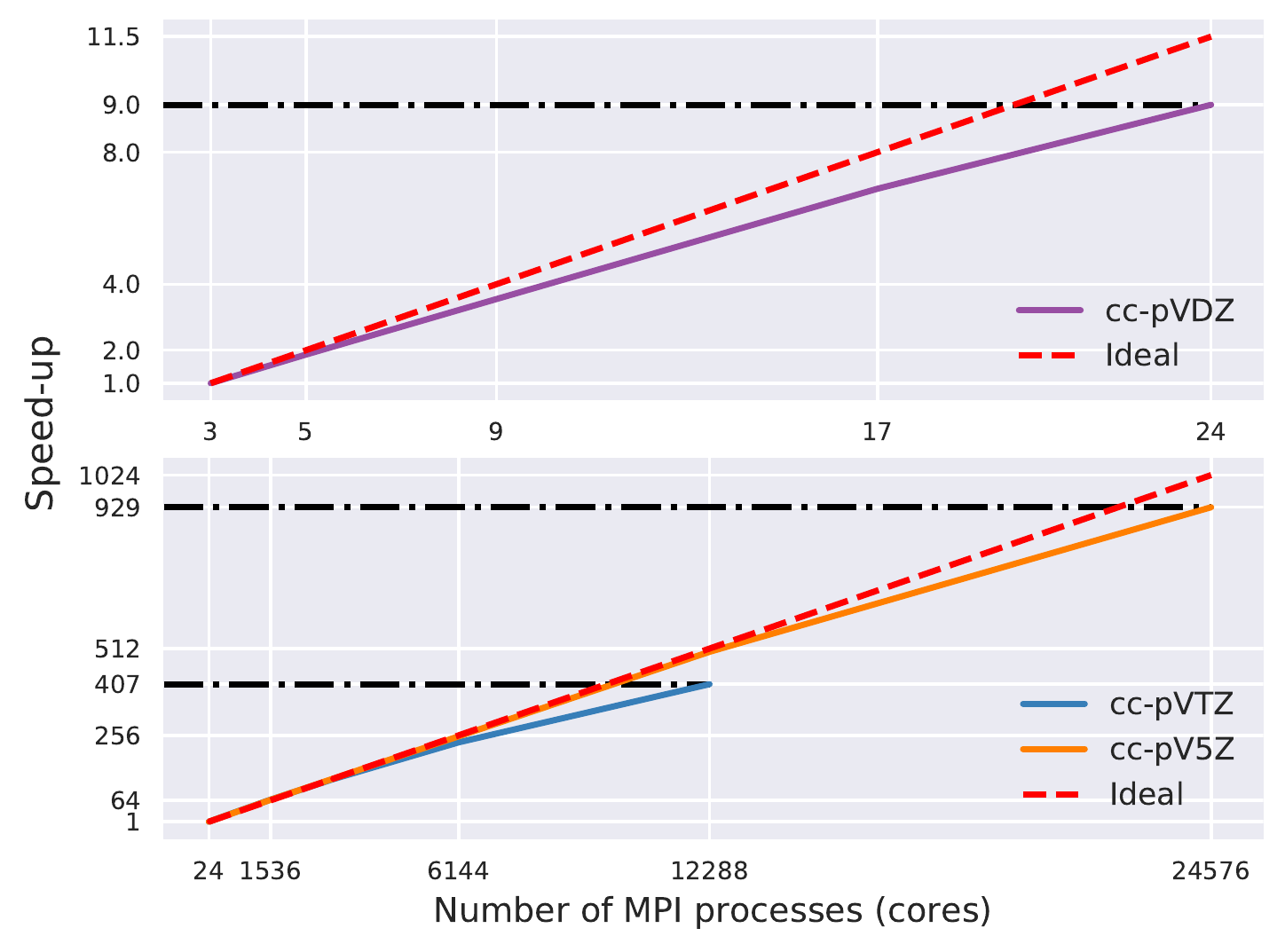}
\caption{Intra- and internode strong scaling for the MBE-FCI-{\textbf{b}} calculations on H$_2$O in Table \ref{h2o_equil_table}.}
\label{h2o_parallel_scaling_fig}
\end{center}
\end{figure}
We conclude this Section with a few results concerned with the parallel scaling potential of the MBE-FCI method. As the method, and hence its implementation within the {\textsc{pymbe}} code, is CPU- rather than memory-bound, strong scalability will mark the only proper measurement of resource utilization. In the upper panel of Figure \ref{h2o_parallel_scaling_fig}, we show the relative speed-up gained by moving from a minimum of $2$ slaves ($3$ cores) to a total of $23$ slaves ($24$ cores) on a single Hazel Hen node (cf. Section \ref{comp_section}) for the MBE-FCI-{\textbf{b}}/cc-pVDZ calculation of Table \ref{h2o_equil_table}. For the intranode parallelization, MPI processes were pinned to individual cores and hyperthreading was disabled. Given that the Sandy Bridge microarchitecture on Hazel Hen does not offer support for more modern features such as fused multiply-add nor the AVX-512 instruction set, it is reasonable to expect the intranode efficiency ($83\%$) in Figure \ref{h2o_parallel_scaling_fig} to be improved on even newer hardware.\\

In terms of internode scalability, the lower panel of Figure \ref{h2o_parallel_scaling_fig} shows the relative speed-up gained by moving from a single Hazel Hen node with MPI employed across all of its 24 cores to a total of $512$/$1024$ nodes (i.e., $12288$/$24576$ individual MPI processes). This time, the scalability has been assessed for the corresponding MBE-FCI-{\textbf{b}} calculations in a medium-sized (cc-pVTZ) and an extended (cc-pV5Z) basis set. We note here that the individual CASCI calculations involving a given number of electrons and orbitals take longer the larger the basis set used is as an integral transformation precedes each of these; this redundancy will be removed in future revisions to the code. At scale, the efficiencies at $512$ ($12288$ cores) and $1024$ nodes ($24576$ cores) amount to $79\%$ and $91\%$ for the expansions in the cc-pVTZ and cc-pV5Z basis sets, respectively. This difference in performance between the two basis sets is mainly ascribed to the larger number of individual CASCI calculations in the latter expansion. In absolute terms, the specific calculations using the cc-pV5Z basis set in Figure \ref{h2o_parallel_scaling_fig} took $608674$ and $655$ seconds on $1$ and $1024$ nodes, respectively. In summary, the MBE-FCI method is thus seen to offer a highly scalable treatment of the electron correlation problem with a massive parallelism ideally suitable for modern supercomputers.

%
%

\section{Summary and Conclusions}\label{conclusion_section}

In the present work, we have extended the recently proposed MBE-FCI method to multideterminantal expansion references in order to make the method applicable to challenging chemical problems dominated by strong electron correlation. Through calculations of the potential energy curves of H$_2$O, N$_2$, C$_2$, and a linear H$_{10}$ chain, this feature enhancement is shown to allow for efficient MBE-FCI calculations that proceed through focussed expansions starting from small compact reference spaces. By comparing the results of the MBE-FCI method to those of a suite of alternative methods, even the use of simple valence space expansion references is shown to enable high accuracy for chemical problems which the standard MBE-FCI method fails to describe satisfactorily. Furthermore, we show that multideterminantal expansion references may be used to compress the involved expansions of the FCI correlation energy to such an extent that near-exact results for H$_2$O and C$_2$ in large basis sets may be obtained even on commodity hardware. For the latter calculations, however, we have also provided numerical results that demonstrate near-ideal parallel scaling of the MBE-FCI method on up to almost 25000 processing units.\\

The results of the present work hence provide further evidence of the fact that the present incremental approach to electron correlation may allow for near-exact calculations to be performed in an unbiased, accurate, and accelerated fashion. However, as is clear from the calculations on stretched C$_2$ and H$_{10}$, the choice of an optimal expansion reference and an optimal set of MOs as the involved expansion objects is yet to be standardized. Indeed, the choice of optimal expansion reference space may differ from what defines a proper reference in standard post-CASSCF methods and may further change in the course of a bond stretch or along a reaction coordinate. In the weakly correlated regime, canonical (or natural) orbitals might offer the most favourable choice, whereas in the strongly correlated regime, some other choice might prove superior (e.g., CASSCF or localized MOs). What defines an optimal expansion reference in the context of MBE-FCI thus warrants more investigation.

%
%
\section*{Acknowledgments}

We are grateful to John F. Stanton of the University of Florida and Devin A. Matthews of the Southern Methodist University for help with some of our CCSDTQ calculations. J. J. E. furthermore thanks Qiming Sun, formerly of California Institute of Technology, as well as the rest of the developer team for their general work on the {\textsc{pyscf}} code. J. J. E. is grateful to the Alexander von Humboldt Foundation for financial support. Finally, we acknowledge PRACE for preparatory access to Hazel Hen at GCS{@}HLRS, Germany.

%
%
\section*{Supporting Information}

All of the results of Figures \ref{h2o_dz_res_fig}, \ref{n2_dz_res_fig}, \ref{c2_dz_res_fig}, and \ref{h10_chain_dz_res_fig} are collected in the Supporting Information as Tables S1--S9.

\newpage

\providecommand*\mcitethebibliography{\thebibliography}
\csname @ifundefined\endcsname{endmcitethebibliography}
  {\let\endmcitethebibliography\endthebibliography}{}


\begin{mcitethebibliography}{82}
\bibitem[Not()]{Note-1}
The H$_2$O/cc-pV5Z calculation using MBE-FCI starting from an HF expansion
  reference was not performed due to time limitations.\relax
\mciteBstWouldAddEndPunctfalse
\mciteSetBstMidEndSepPunct{\mcitedefaultmidpunct}
{}{\mcitedefaultseppunct}\relax
\EndOfBibitem
\end{mcitethebibliography}
\end{document}